\documentclass[3p]{elsarticle}
\usepackage{lineno,hyperref}
\usepackage{subcaption}
\usepackage{amsmath}
\usepackage{forest}
\usepackage{courier}
\usetikzlibrary{shapes,arrows,positioning,automata}

\tikzset{
  treenode/.style = {shape=rectangle, rounded corners,
                     draw, align=center,
                     top color=white,
                     bottom color=white},
  root/.style     = {treenode, font=\Large,
                     bottom color=red!30},
  env/.style      = {treenode, font=\ttfamily\normalsize},
  dummy/.style    = {circle,draw}
}
\usepackage{caption}
\usepackage{listings}
\usepackage{xcolor} 
\usepackage{threeparttable}
\usepackage{algorithm,algpseudocode}
\usepackage[usestackEOL]{stackengine}
\usepackage{xcolor}
\usepackage[english]{babel}
\usepackage[utf8]{inputenc}

\hypersetup{
  colorlinks=true,
  urlcolor={blue!80!black}
}

\edef\tmp{\the\baselineskip}
\setstackgap{L}{\tmp}
\newcommand{\myref}[1]{A}

\journal{Journal} 

\begin{document}

\begin{frontmatter}

\title{Reliability Estimation of an Advanced Nuclear Fuel using Coupled Active Learning, Multifidelity Modeling, and Subset Simulation}




\author[A1]{Somayajulu L. N. Dhulipala\corref{mycorrespondingauthor}}
\address[A1]{Computational Mechanics and Materials, Idaho National Laboratory, Idaho Falls, ID 83402, USA}\cortext[mycorrespondingauthor]{Corresponding author; Email: Som.Dhulipala@inl.gov}
\author[A2]{Michael D. Shields}
\author[A2]{Promit Chakroborty}
\address[A2]{Department of Civil and Systems Engineering, Johns Hopkins University, Baltimore, MD 21218, USA}
\author[A1]{Wen Jiang}
\author[A1]{Benjamin W. Spencer}
\author[A1]{Jason D. Hales}
\author[A5]{Vincent M. Labour\'e}
\address[A5]{Reactor Physics Methods and Analysis, Idaho National Laboratory, Idaho Falls, ID 83402, USA}
\author[A5]{Zachary M. Prince}
\author[A3]{Chandrakanth Bolisetti}
\address[A3]{Advanced Reactor Technology and Design, Idaho National Laboratory, Idaho Falls, ID 83402, USA}
\address[A4]{Computational Frameworks, Idaho National Laboratory, Idaho Falls, ID 83402, USA}
\author[A1]{Yifeng Che}


\begin{abstract}
Tristructural isotropic (TRISO)-coated particle fuel is a robust nuclear fuel and determining its reliability is critical for the success of advanced nuclear technologies. However, TRISO failure probabilities are small and the associated computational models are expensive. We used coupled active learning, multifidelity modeling, and subset simulation to estimate the failure probabilities of TRISO fuels using several 1D and 2D models. With multifidelity modeling, we replaced expensive high-fidelity (HF) model evaluations with information fusion from two low-fidelity (LF) models. For the 1D TRISO models, we considered three multifidelity modeling strategies: only Kriging, Kriging LF prediction plus Kriging correction, and deep neural network (DNN) LF prediction plus Kriging correction. While the results across these multifidelity modeling strategies compared satisfactorily, strategies employing information fusion from two LF models consistently called the HF model least often. Next, for the 2D TRISO model, we considered two multifidelity modeling strategies: DNN LF prediction plus Kriging correction (data-driven) and 1D TRISO LF prediction plus Kriging correction (physics-based). The physics-based strategy, as expected, consistently required the fewest calls to the HF model. However, the data-driven strategy had a lower overall simulation time since the DNN predictions are instantaneous, and the 1D TRISO model requires a non-negligible simulation time.
\end{abstract}






\end{frontmatter}


\section{Introduction}

Tristructural isotropic (TRISO)-coated particle fuel is an advanced nuclear fuel designed to withstand extreme operating temperatures inside reactors \cite{Maki2007a,Hales2013a,Gamble2021a}. Due to its robustness, it is being proposed to be used for multiple modern nuclear technologies, such as small modular reactors, advanced reactors, and microreactors \cite{DOE_TRISO}. TRISO particles can be compacted in multiple forms for various applications. A $25$-mm-long, $6$-mm-radius cylindrical fuel compact can contain approximately 10,000--15,000 TRISO particles, each with a radius of around $375-430$ $\mu$m \cite{Petti2012a}. Each TRISO particle has a fuel kernel and several protective outer layers, and a particle is considered to have failed if the outer protective layer has fractured due to thermo-mechanical stresses \cite{Jiang2021a}. Due to the uncertainties in the fuel properties and the particle geometries, the failure of a TRISO particle is characterized probabilistically, and the failure probabilities can range from $10^{-3}$ to $10^{-6}$ depending upon irradiation conditions in the reactor and fuel fabrication processes \cite{Jiang2021a,Miller2018a}. Determining these failure probabilities accurately is critical to assess the ability to retain fission products within the fuel \cite{Hales2021a,Jiang2021a,Miller2018a} and further analyze the overall reactor safety \cite{Paaren2021a}. However, TRISO models can be computationally expensive, which makes it difficult to compute these failure probabilities using regular Monte Carlo or even standard variance reduction methods, such as subset simulation \cite{Au2001a}. Active learning for reliability estimation provides a good alternative to efficiently estimate small failure probabilities involving expensive computational models (e.g., \cite{Razaaly2018a,Razaaly2020a,Cui2019a,Dhulipala_AL_MFM}). In addition, there has been an interest in multifidelity modeling strategies for uncertainty quantification that fuse data from low-fidelity (LF) and high-fidelity (HF) models (e.g., \cite{Peherstorfer2018a,Kramer2019a,Pham2021a,Gorodetsky2020a}). Combining active learning and multifidelity modeling can further reduce the computational expense when estimating low failure probabilities. This study applies a coupled active learning, multifidelity modeling, and subset simulation algorithm proposed in \citet{Dhulipala_AL_MFM} to estimate the failure probabilities and reliability indices (i.e., negative of the inverse CDF corresponding to a standard Normal distribution of the failure probabilities) of TRISO fuel models.

Active learning methods for efficient small failure probability estimation are popular when the computational model is expensive. These active learning methods leverage machine-learned surrogate models to efficiently approximate the HF model and drive the learning, most often using Gaussian Process regression (Kriging; \cite{Che2021a}) or polynomial chaos expansions \cite{Wu2021a}.  The Adaptive Kriging Monte Carlo Simulation (AK-MCS) method proposed by \citet{Echard2011a} is a popular approach for efficient reliability estimation. As such, several versions of the AK-MCS algorithm have been developed for complex limit state functions and/or high-dimensional spaces \cite{Lelievre2018a,Haj2021a,Ameryan_SSC}. There has also been interest in integrating variance reduction methods, such as importance sampling and subset simulation, with adaptive Kriging for reduced computational cost along with accurate failure probability estimation \cite{Huang2016a,Zhang2019a,Xu2020a}. In addition to the use of Kriging as the primary surrogate to replace expensive high-fidelity (HF) model evaluations, studies such as \citet{Papadopoulos_NN} and \citet{Cheng_SVM} use other machine learning models, such as support vector machines and neural networks, for reliability estimation. However, most active learning methods for reliability estimation rely on a single surrogate model and do not incorporate multifidelity physics-based models. Independently of the problem of small failure probability estimation, the concept of multifidelity modeling is being utilized, where information from multiple LF models are fused to approximate the HF model output \cite{Peherstorfer2016y,Ranftl2020a}. Studies have shown that, for many uncertainty quantification tasks, combining information from multiple LF models can reduce the computational expense compared to a single LF model \cite{Peherstorfer2018a,Perdikaris2017a,Gorodetsky2020a}. Based on this concept, \citet{Dhulipala_AL_MFM} proposed a coupled active learning, multifidelity modeling, and subset simulation algorithm for efficient low-failure probability computations. Their algorithm adds a Kriging-based correction term to an LF model output to adaptively decide when to call a HF model. This algorithm has some interesting features such as: (1) flexibility over the LF model selection and potentially to allow for multiple LF models; (2) the capability to estimate very low failure probabilities on the order of $10^{-9}$ (most, if not all, active learning algorithms can have difficulty estimating such very low failure probabilities as noted in \cite{Razaaly2018a,Razaaly2020a,Cui2019a}); and (3) minimal memory and computational requirements for Kriging usage as the test size is always one (the computational complexity of Kriging for prediction and uncertainty quantification with input dimension $d$ and test size $S$ are $\mathcal{O}(dS)$ and $\mathcal{O}(S^2)$, respectively \cite{Raykar2007}).

In this work, we apply the coupled active learning, multifidelity modeling, and subset simulation algorithm proposed in \cite{Dhulipala_AL_MFM} for failure probability estimation of multifidelity TRISO fuel models, considering both 1D and 2D versions of TRISO fuel models. While the 1D model takes about 11 seconds for each evaluation, the 2D TRISO model takes about 4 minutes on ten processors (this also depends on the mesh density) and {can characterize the TRISO failure more accurately}. TRISO models can have failure probabilities from $10^{-3}$ to $10^{-6}$. As such, contributions of this paper are as follows:
\begin{itemize}
    \item This study illustrates the computational gains brought by active learning on practical and critical problems regarding the safety of advanced nuclear fuel systems, while the majority of studies proposing active learning methods are demonstrated on analytical limit state functions or HF models of academic interest.
    \item This paper explores whether multiple LF models in active learning can bring computational gains as opposed to a single LF model (e.g., a Kriging model) that is routinely used.
    \item Data-driven and physics-based multifidelity modeling strategies are considered and the computational gains within and across these strategies are discussed for 1D and 2D TRISO models failure estimation.
\end{itemize}


This paper is organized as follows. Section \ref{TRISO_desc} gives an introduction to TRISO fuel modeling. Section \ref{sec:AL_background} provides some background on Kriging, the AK-MCS algorithm, and subset simulation. Section \ref{sec:AL_MFM} summarizes the coupled active learning, multifidelity modeling, and subset simulation algorithm proposed in \citet{Dhulipala_AL_MFM}. Section \ref{sec:1D_TRISO} applies this coupled algorithm for the reliability estimation of TRISO fuel using 1D models as the HF model by considering three multifidelity modeling strategies. Section \ref{sec:2D_TRISO} applies the coupled algorithm for the reliability estimation of TRISO fuel using 2D models as the HF model by considering two multifidelity modeling strategies, which are data-driven and physics-based. Section \ref{sec:sum_conc} summarizes the contributions of this paper and presents the important conclusions.

\section{Description of the TRISO Nuclear Fuel Model}\label{TRISO_desc}

\begin{figure}[H]
\centering
\includegraphics[width=2.75in, height=1.222in]{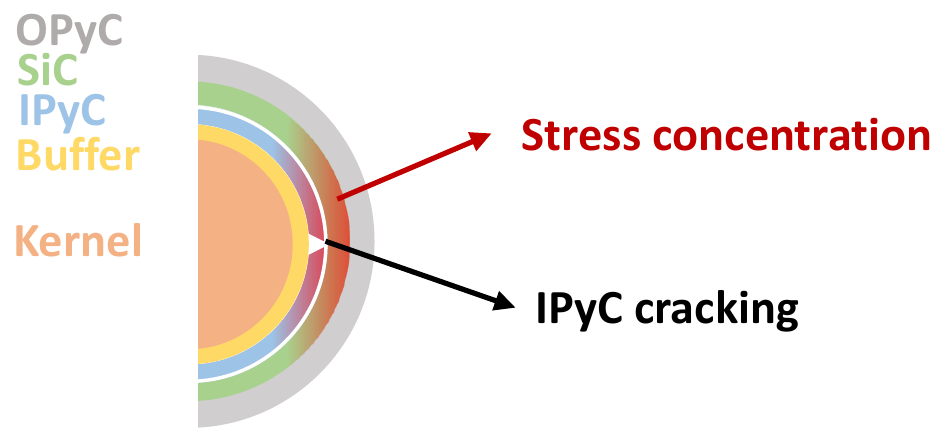}
\caption{Axisymmetric representation of a TRISO particle without asphericity.}
\label{TRISO}
\end{figure}

The TRISO particle is composed of a fuel kernel and four layers: buffer, inner pyrolytic carbon (IPyC), silicon carbide (SiC), and outer pyrolytic carbon (OPyC). Figure \ref{TRISO} presents a schematic of an axisymmetric representation of the TRISO particle. Each TRISO particle has a radius of about $430~\mu$m. Due to fission, the fuel kernel generates heat, which is transferred to the coating layers, resulting in thermo-mechanical stresses, which increase susceptibility to failure. IPyC and SiC layers retain fission products within the TRISO particles, therefore, maintaining the structural integrity of these layers is particularly important. The TRISO particle is considered to have failed when the SiC layer fails, which means the stress exceeds its strength. Two failure modes for SiC failure are considered \cite{Jiang2021a,Dhulipala_TRISO}: stress concentration due to IPyC cracking and pressure vessel failure. When the stress in the IPyC layer exceeds its strength, it cracks and causes a stress concentration in the SiC layer (Figure \ref{TRISO}). The resulting stress concentration increases the likelihood of SiC layer failure. Alternatively, the SiC layer can fail without IPyC layer cracking when its stress exceeds its strength. This failure mode is called a pressure vessel failure and is more likely in an aspherical particle than in the spherical one presented in Figure \ref{TRISO}. However, a SiC layer failure in the pressure vessel mode is 3--4 orders of magnitude less likely than a SiC failure due to IPyC cracking in both spherical and aspherical particles (see the UCO kernel discussion in \citet{Jiang2021a} and \citet{Miller2018a}). As such, only the failure due to IPyC cracking is considered and the failure probability is estimated using:

\begin{equation}
    \label{eqn:TRISO_Pf}
    P_f = P(\textrm{SiC failure}) = P(\textrm{SiC failure}~|~\textrm{IPyC failure})~P(\textrm{IPyC failure})
\end{equation}

\noindent Because TRISO fuel failure depends heavily on the local mechanical and thermal response, accurate models for that are an essential foundation for failure prediction. The governing equations for modeling the thermo-mechanical behavior of TRISO, levels of modeling fidelity, and model input parameters are discussed in the following sections.

\subsection{Governing Equations}

The thermo-mechanical behavior of the fuel kernel and each layer of a TRISO particle is governed by two equations: heat and momentum conservation as described in \citet{Williamson2012a} and \citet{Jiang2021a}. The heat equation is given by:

\begin{equation}
    \label{eqn:GE_Heat}
    \rho~{C}_p~\frac{\partial T}{\partial t} + \nabla \pmb{\cdot} (-k \nabla T) - {E}_f~{\dot{F}} = 0
\end{equation}

\noindent where $T$ is the unknown temperature and $k$, $C_p$, $\rho$, $E_f$, and ${\dot{F}}$ are the thermal conductivity, specific heat, density, energy released per fission, and volumetric fission rate, respectively. The momentum conservation equation (neglecting body forces and inertial forces) is given by:

\begin{equation}
    \label{eqn:GE_Momen}
    \nabla \pmb{\cdot} \pmb{\sigma} = 0 
\end{equation}

\noindent where $\pmb{\sigma}$ is the Cauchy stress tensor. The constitutive law (assuming small strains) is given by:

\begin{equation}
    \label{eqn:GE_Constit}
    \pmb{\sigma} = \mathcal{C} : (\pmb{\varepsilon}-\pmb{\varepsilon}_c-\pmb{\varepsilon}_t-\pmb{\varepsilon}_i)
\end{equation}

\noindent where $\mathcal{C}$ is the elasticity tensor and $\pmb{\varepsilon}$, $\pmb{\varepsilon}_c$, $\pmb{\varepsilon}_t$, and $\pmb{\varepsilon}_i$ are the total strain, irradiation creep strain, thermal expansion strain, and irradiation-induced eigen strain, respectively. These governing equations for the fuel kernel and outer layers are dependent on a number of parameters and coefficients calibrated from experiments. \citet{Jiang2021a} provides an exhaustive description of the governing equations and their parameters and coefficients for different TRISO layers.

\subsection{Multifidelity TRISO Fuel Models and Failure Indicator}

\begin{figure}[H]
\centering
\includegraphics[width=4.5in, height=2.0in]{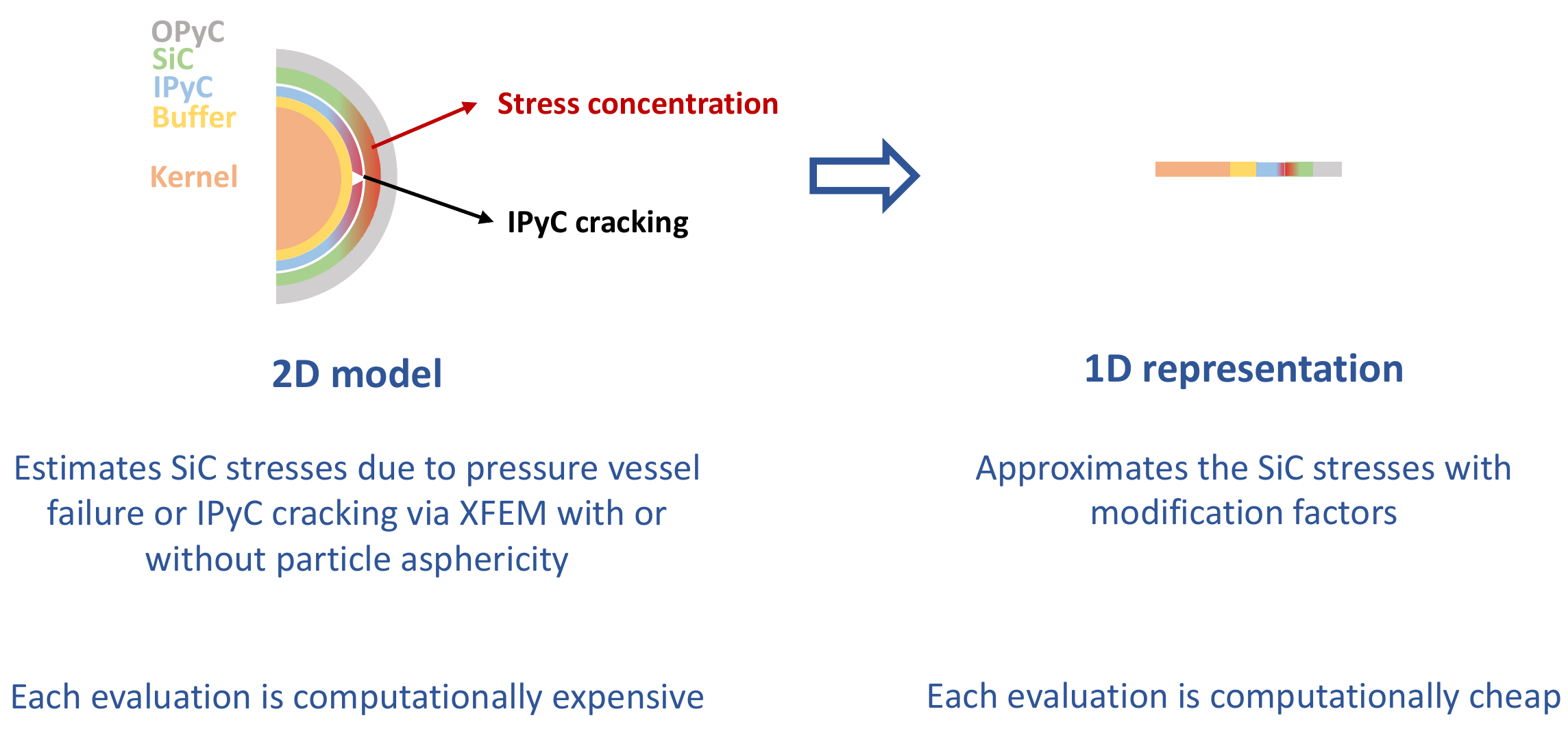}
\caption{2D and 1D representations of the TRISO model along with details for calculating thermo-mechanical stresses in the protective layers. This figure is adapted from \citet{Dhulipala_TRISO}.}
\label{TRISO_failure}
\end{figure}

Multifidelity modeling of the TRISO particle is performed here using the Bison fuel performance code \cite{Williamson2021a}, which is built using the Multiphysics Object Oriented Simulation Environment \cite{Permann2020a} \newline (MOOSE; \href{https://mooseframework.inl.gov}{https://mooseframework.inl.gov}). To model IPyC cracking and stress concentration in SiC as presented in Figure \ref{TRISO}, a 2D model of the TRISO particle is required. Such a 2D model relies on the extended finite element method (XFEM) module in MOOSE \cite{Jiang2020a} to represent a crack in the IPyC layer and simulate stress concentration in the SiC layer. Depending upon the mesh density in the finite element model, each evaluation of the 2D TRISO model can take about {4 minutes} on ten processors, which can be too expensive for probabilistic analysis. Therefore, a 1D version of the TRISO model has been developed for efficient failure analysis \cite{Miller2018a,Jiang2021a}, as shown in Figure \ref{TRISO_failure}. The 1D model cannot directly model cracking of the IPyC layer and the resulting stress concentration in the SiC layer. Stress modification factors are used in the 1D model to approximate the stress concentrations in a corresponding 2D model \cite{Jiang2021a}. The 1D model, therefore, approximates the behavior of the 2D model, although it is much more computationally efficient. Each 1D model evaluation takes about 11 seconds.

For the 2D and 1D versions of the TRISO model, either IPyC or SiC layer failure is defined by the indicator function:

\begin{equation}
    \label{TRISO_Ind}
    \mathbf{I} = \begin{cases}
    1~~~~~\textrm{if}~~\sigma_1(\pmb{X}^\#)-\Sigma \geq 0\\
    0~~~~~\textrm{if}~~\sigma_1(\pmb{X}^\#)-\Sigma < 0
    \end{cases}
\end{equation}

\noindent where $\sigma_1$ is the maximum principal stress in either IPyC or SiC, $\pmb{X}^\#$ is a subset of the input parameters, and $\Sigma$ is either IPyC or SiC strength. The full set of input parameters include the IPyC and SiC strengths and is given by $\pmb{X} = \{\pmb{X}^\#,~\Sigma^{\textrm{IPyC}},~\Sigma^{\textrm{SiC}}\}$. The whole set of uncertain input parameters $\pmb{X}$ to the TRISO model is discussed next.

\subsection{Considered TRISO Models and Input Parameter Uncertainties}\label{TRISO_unc}

The TRISO model is defined by a number of input parameters, of which the layer geometries and mechanical properties can be uncertain and have been found to be the most important for its probabilistic failure analysis. We considered four TRISO models in this study to determine their failure probabilities and these models represent a range of verification cases as well as experimental validation cases as discussed in \citet{Jiang2021a}. These four models differ not only in terms of their distributions of layer geometries and mechanical properties but also in the irradiation temperatures inside the nuclear reactor. Both the input parameter distributions and the irradiation temperatures can affect the failure probabilities. Tables \ref{Table_input_params} and \ref{Table_input_temp} present the input parameters and irradiation temperatures for the four TRISO models, respectively. While Models 1 \& 2 assume that the TRISO particle is perfectly spherical, Models 3 \& 4 eliminate this assumption by using an asphericity ratio. In addition, for Models 1 \& 2, the irradiation temperature is considered transient during the simulation, and the maximum and minimum values are presented in Table \ref{Table_input_temp}; please refer to \citet{Jiang2021a} for a discussion on how the irradiation temperature changes with time. For Models 3 \& 4, the irradiation temperature is constant. In summary, Models 1 \& 2 and Models 3 \& 4 have 7 and 11 uncertain input parameters, respectively.

\begin{table}[H]
\centering
\caption{Input parameters for the four TRISO models considered in this study. This table is adapted from \citet{Dhulipala_TRISO}.}
\label{Table_input_params}
\small
\begin{tabular}{ |c|c|c|c| }
\hline
\Centerstack[c]{\textbf{Category}} & \Centerstack[c]{\textbf{Parameter}} & \Centerstack[c]{\textbf{Models 1 \& 2}} & \Centerstack[c]{\textbf{Models 3 \& 4}}\\
\hline
\Centerstack[c]{Particle \\ geometry} & \Centerstack[c]{Kernel radius ($\mu m$) \\ Buffer thickness ($\mu m$) \\ IPyC thickness ($\mu m$) \\ SiC thickness ($\mu m$) \\ OPyC thickness ($\mu m$) \\ Asphericity ratio} & \Centerstack[c]{$\mathcal{N}(213.35,~4.4)$ \\ $\mathcal{N}(98.9,~8.4)$ \\ $\mathcal{N}(40.4,~2.5)$ \\ $\mathcal{N}(35.2,~1.2)$ \\ $\mathcal{N}(43.4,~2.9)$ \\ $1.0$} & \Centerstack[c]{$\mathcal{N}(212.5,~5.0)$ \\ $\mathcal{N}(100.0,~10.0)$ \\ $\mathcal{N}(40.0,~3.0)$ \\ $\mathcal{N}(35.0,~2.0)$ \\ $\mathcal{N}(40.0,~3.0)$ \\ $1.04$}\\
\hline
\Centerstack[c]{Fuel \\ properties} & \Centerstack[c]{Kernel density ($gm/cm^3$) \\ Kernel theoretical density ($gm/cm^3$) \\ Buffer density ($gm/cm^3$) \\ Buffer theoretical density ($gm/cm^3$) \\ IPyC density ($gm/cm^3$) \\ OPyC density ($gm/cm^3$) \\ IPyC anisotropy factor \\ OPyC anisotropy factor} & \Centerstack[c]{10.966 \\ 11.37 \\ 1.05 \\ 2.25 \\ 1.89 \\ 1.907 \\ 1.0465 \\ 1.0429} & \Centerstack[c]{11.0 \\ 11.4 \\ 1.05 \\ 2.25 \\ $\mathcal{N}(1.9,~0.02)$ \\ $\mathcal{N}(1.9,~0.02)$ \\ $\mathcal{N}(1.05,~0.005)$ \\ $\mathcal{N}(1.05,~0.005)$}\\
\hline
\Centerstack[c]{Layer \\ strengths} & \Centerstack[c]{IPyC strength \\ SiC strength} & \Centerstack[c]{$\mathcal{W}(\sigma_{ms},~9.5)$ \\ $\mathcal{W}(\sigma_{ms},~6.0)$} & \Centerstack[c]{$\mathcal{W}(\sigma_{ms},~9.5)$ \\ $\mathcal{W}(\sigma_{ms},~6.0)$}\\
\hline
\end{tabular}
\begin{tablenotes}
\item[] \textbf{Notations:} $\mathcal{N}(x,~y)$: normal distribution with mean $x$ and standard deviation $y$; $\sigma_{ms}$: mean strength, which is dependent on the characteristic strength obtained from \citet{Jiang2021a}; $\mathcal{W}(\sigma_{ms},~m)$: Weibull distribution with mean strength $\sigma_{ms}$ and Weibull modulus $m$.
\end{tablenotes}
\end{table}

\begin{table}[H]
\centering
\caption{Irradiation temperatures in the nuclear reactor for the four TRISO fuel models.}
\label{Table_input_temp}
\small
\begin{tabular}{ |c|c|c|c| }
\hline
\Centerstack[c]{\textbf{Model 1}} & \Centerstack[c]{\textbf{Model 2}} & \Centerstack[c]{\textbf{Model 3}} & \Centerstack[c]{\textbf{Model 4}}\\
\hline
\Centerstack[c]{Type = Daily varying \\ Max. = $1226.84^o$C \\ Min. = $207.4^o$C} & \Centerstack[c]{Type = Daily varying \\ Max. = $1281.84^o$C \\ Min. = $195.84^o$C} & \Centerstack[c]{Type = Constant \\ Value = $700.0^o$C} & \Centerstack[c]{Type = Constant \\ Value = $1000.0^o$C}\\
\hline
\end{tabular}
\end{table}

\section{Background on Kriging, Active Learning in Monte Carlo Sampling, and Subset Simulation}\label{sec:AL_background} 

This section provides a brief background on Kriging, active learning in Monte Carlo sampling, and subset simulation. These concepts provide a basis for the coupled active learning, subset simulation, and multifidelity modeling discussed in the next section.

\subsection{Kriging with Multiple Length Scales}

The working principle of Kriging has been described well in several studies related to active learning \cite{Echard2011a,Dhulipala_AL_MFM}. Only a brief overview will be provided here for completeness. In Kriging, given the training set $(\pmb{X},~\pmb{y})$ and test inputs $\pmb{X}_*$, we are interested in inferring the posterior predictive distribution of $\pmb{y}_*$, which has a closed form solution \cite{Rasmussen2004}:

\begin{equation}
    \label{eqn:Kriging_1}
    \begin{aligned}
    p(\pmb{y}_*~|~\pmb{X}, \pmb{X}_*, \pmb{y}) \sim \mathcal{N}\Big(~ & k(\pmb{X}_*,\pmb{X})~k(\pmb{X},\pmb{X})^{-1}~\pmb{y}, \\
    & k(\pmb{X}_*,\pmb{X}_*) - k(\pmb{X}_*,\pmb{X})~k(\pmb{X},\pmb{X})^{-1}~k(\pmb{X},\pmb{X}_*)~\Big)
    \end{aligned}
\end{equation}

\noindent where $k(.,.)$ is the kernel function that describes the correlation between two points in the input space and $\mathcal{N}(.,.)$ is the multivariate Normal distribution. Although the posterior predictive distribution has a closed form solution in a Kriging model, the kernel function has hyperparameters, which need to be optimized given the training data. This optimization is performed by maximizing the negative log likelihood function:

\begin{equation}
    \label{eqn:Kriging_2}
    \mathcal{L} = -\ln~p(\pmb{y}~|~\pmb{X},\sigma^2,\lambda) \propto \frac{1}{2}~\ln |k(\pmb{X},\pmb{X})| + \frac{1}{2}~\pmb{y}^T~k(\pmb{X},\pmb{X})^{-1}~\pmb{y}
\end{equation}

\noindent The kernel function in Equations \eqref{eqn:Kriging_1} and \eqref{eqn:Kriging_2} is usually described by a squared exponential kernel \cite{Duvenaud2014a}:

\begin{equation}
    \label{eqn:Kriging_3}
    k(\pmb{X},\pmb{X}_*) = \sigma^2~\exp \Bigg(-\frac{(\pmb{X}-\pmb{X}_*)^2}{2l^2}\Bigg)
\end{equation}

\noindent where $\sigma^2$ is the amplitude scale and $l$ is the length scale. The length scale, in general, describes the degree with which the outputs change with respect to the inputs. In practice, it is possible for different input parameters in the vector $\pmb{X}$ to have different degrees of sensitivity on the output. In such cases, having a single length scale as in Equation \eqref{eqn:Kriging_3} is not advisable. Using the concept of anisotropic kernels, the squared exponential kernel can be modified to have multiple length scales, one for each input parameter in the vector $\pmb{X}$ \cite{Duvenaud2014a}:

\begin{equation}
    \label{eqn:Kriging_4}
    k(\pmb{X},\pmb{X}_*) = \sigma^2~\exp \Bigg(-\frac{1}{2}\sum_{d=1}^D\frac{(X_d-{X}_{*d})^2}{l_d^2}\Bigg)
\end{equation}

\noindent where $l_d$ is the length scale corresponding to dimension $d$ in the input vector $\pmb{X}$ and $D$ is the dimensionality of $\pmb{X}$.

\subsection{Active Learning in Monte Carlo Sampling: AK-MCS Method}

In a regular Monte Carlo sampling (MCS), an estimator for the failure probability is given by:

\begin{equation}
    \label{eqn:AKMCS_1}
    \hat{P}_f = \frac{1}{N}\sum_{i=1}^{N}~I\bigg(F(\pmb{X}_i)\leq 0\bigg) = \frac{1}{N}\sum_{i=1}^{N}~\mathbf{I}_{{{i}}}
\end{equation}

\noindent where $N$ is number of Monte Carlo samples, $F(\pmb{X}_i)\leq 0$ is the condition that the HF model output for $i^{\textrm{th}}$ input parameters sample $\pmb{X}_i$ exceeds the specified failure threshold, and $\mathbf{I}_i$ is an indicator function for model output exceeding the failure threshold. Reliability estimation often involves low failure probabilities, and estimating these probabilities requires numerous calls to a computationally expensive HF model. Owing to the drawbacks from a computational perspective of using a regular MCS, \citet{Echard2011a} proposed an adaptive use of Kriging to replace expensive HF model evaluations with cheaper Kriging predictions. With the use of adaptive Kriging in MCS (AK-MCS), the estimator for the failure probability becomes:

\begin{equation}
    \label{eqn:AKMCS_2}
    \hat{P}_f = \frac{1}{N}\sum_{i=1}^{N}~\mathcal{P}_i(\pmb{X}) = \frac{1}{N}\sum_{i=1}^{N}~\mathcal{P}_i
\end{equation}

\noindent where $\mathcal{P}_i$ is the probability that $\mathbf{I}_{{{i}}}=1$ and is further expanded as:

\begin{equation}
    \label{eqn:AKMCS_3}
    \mathcal{P}_i = P(\mathbf{I}_{{{i}}}=1) = P(\mathbf{I}_{i}=1 | \mathbf{I}_{i, {\mathcal{K}}} = 1)~P( \mathbf{I}_{i, {\mathcal{K}}} = 1) + P(\mathbf{I}_{i}=1 | \mathbf{I}_{i, {\mathcal{K}}} = 0)~P( \mathbf{I}_{i, {\mathcal{K}}} = 0)
\end{equation}

\noindent where $\mathbf{I}_{i, {\mathcal{K}}}$ is the value of the indicator function when Kriging predictions are used. Equation \eqref{eqn:AKMCS_3} takes into consideration the probability of Kriging predictions making a sign error (i.e., falsely classifying a failed sample as being safe and vice versa) through the use of conditional probabilities $P(\mathbf{I}_{i}=1 | .)$. However, since the accuracy of Kriging predictions is checked for every sample, $\mathcal{P}_i$ in Equation \eqref{eqn:AKMCS_2} is expressed as:

\begin{equation}
    \label{eqn:AKMCS_4}
    \mathcal{P}_i = \begin{cases} \begin{cases}
1 \times {\Phi}_{{{i}}} + 0 \times (1-{\Phi}_{{{i}}}) = {\Phi}_{{{i}}}~~~~~~~~~~\text{if $\mathbf{I}_{{{i}},\mathcal{K}} = 1$} \\
0 \times {\Phi}_{{{i}}} + 1 \times (1-{\Phi}_{{{i}}})  = 1-{\Phi}_{{{i}}}~~~~~\text{if $\mathbf{I}_{{{i}},\mathcal{K}} = 0$}
\end{cases} \text{for Kriging prediction} \\ ~~\mathbf{I}_{i} ~~~~~~~~~~~~~~~~~~~~~~~~~~~~~~~~~~~~~~~~~~~~~~~~~~~~~~~~~~~~~\text{for HF model prediction} \end{cases}
\end{equation}

\noindent where ${\Phi}_{{{i}}}$ is the probability of a Kriging prediction making a sign error. To determine when to call the HF model and when a Kriging prediction is used in Equation \eqref{eqn:AKMCS_4}, certain Kriging learning functions are used. Several Kriging learning functions have been developed for use in reliability analysis (e.g., \citet{bichon2008efficient}, \citet{Echard2011a}, or \citet{Zhang2019a}). The $U$-function is a popular one and is defined as \cite{Echard2011a}:

\begin{equation}
    \label{eqn:AKMCS_5}
    U = \frac{|\mu_{{\mathcal{K}}}(\pmb{X})|}{\sigma_{{\mathcal{K}}}(\pmb{X})}
\end{equation}

\noindent where $\mu_{{\mathcal{K}}}(\pmb{X})$ is the mean Kriging prediction and $\sigma_{{\mathcal{K}}}(\pmb{X})$ is its standard deviation. The $U$-function becomes sensitive to the accuracy of Kriging predictions for points that are close to the failure threshold and/or associated with a high standard deviation. New points are iteratively added to the training set, and the HF model is evaluated at training points that correspond to the minimum of $U$. The Kriging model is re-trained, and this process is repeated until $\min(U)\ge 2$. This value of 2 corresponds to a probability of making a sign error of $\Phi_i = \Phi(-U_i) = \Phi(-2) \approx 0.0228$ ($\Phi$ is a standard Normal cumulative distribution function). Also, as explained in \citet{Dhulipala_AL_MFM}, for most cases when the $U$-function exceeds $2$ and Kriging predictions are used, $\Phi_i$ in Equation \eqref{eqn:AKMCS_4} will approximately equal $1$. Therefore, $\mathcal{P}_i$ in Equations \eqref{eqn:AKMCS_2} and \eqref{eqn:AKMCS_3} converges to an indicator function $\mathbf{I}_i$. Also, the coefficient of variation (COV) estimator for the failure probability converges to the MCS COV estimator:

\begin{equation}
    \label{eqn:AKMCS_6}
    \hat{\gamma} = \sqrt{\frac{1-\hat{P}_f}{\hat{P}_f~N}}
\end{equation}

\noindent where $N$ is the total number of calls to both the high-fidelity and Kriging models.

A notable drawback of AK-MCS, as explained in \citet{Dhulipala_AL_MFM}, is that when failure probabilities are $10^{-4}$ or less, AK-MCS requires evaluating a large number of Kriging predictions for a satisfactory COV value. A large number of Kriging evaluations can result in computational and memory requirement bottlenecks~\cite{Raykar2007}, and approaches building on AK-MCS have been developed to alleviate these issues as discussed below.


\subsection{Subset Simulation}

The following section presents a brief overview of subset simulation for the sake of completeness; more details can be found in several other studies \cite{Au2001a,Xu2020a,Huang2016a}. Subset simulation expresses the failure probability $\hat{P}_f$ as a product of larger failure probabilities:

\begin{equation}
    \label{eqn:SS_1}
    \hat{P}_f = \hat{P}_1~\prod_{i=2}^{N_s}~\hat{P}_{i|i-1}
\end{equation}

\noindent where $\hat{P}_1$ is the first unconditional failure probability computed as the fraction of samples exceeding the intermediate failure threshold, $\mathcal{F}_1$, and $\hat{P}_{i|i-1}$ are the subsequent conditional failure probabilities that are conditional on exceeding the prior intermediate failure thresholds, $\mathcal{F}_{i-1}$, and are computed as the fraction of samples exceeding the intermediate failure threshold $\mathcal{F}_i$. In expressing $\hat{P}_f$ as a product of larger failure probabilities, subset simulation creates intermediate failure thresholds $(\mathcal{F}_1, \dots, \mathcal{F}_{N_s-1})$ before the required threshold of zero. The samples falling in between two subsequent intermediate failure thresholds $\mathcal{F}_{i-1}$ and $\mathcal{F}_{i}$ constitute a subset. While a regular MCS is used to estimate $\hat{P}_1$, a Markov chain Monte Carlo is used to estimate $\hat{P}_{i|i-1}$. To estimate these intermediate failure probabilities, the intermediate failure thresholds must be specified. These are usually taken as the $1-p_o$ percentile value of sample outputs from each subset. $p_o$ is fixed \emph{a priori} (0.1 is usually used). Although a component-wise Metropolis-Hastings algorithm proposed by \citet{Au2001a} is popular for simulating the conditional samples in $\hat{P}_{i|i-1}$, it can sometimes lead to degenerate sampling when dealing with models having geometrically complex limit state functions. To alleviate these issues, more advanced sampling methods, such as delayed rejection Markov chain Monte Carlo \cite{Papaioannou2015a}, Hamiltonian Monte Carlo \cite{Wang2019a}, and affine invariant ensemble sampling \cite{Shields2021a}, have been proposed. In addition, an approximate COV estimate to the intermediate failure probability in subset $i$ is given by \cite{Au2001a}:

\begin{equation}
    \label{eqn:SS_2}
    \hat{\delta}_i =
    \begin{cases}
    \sqrt{\frac{1-\hat{P}_1}{\hat{P}_1~N_n}}~~~~~\forall~~i=1 \\
    \\
    \sqrt{\frac{1-\hat{P}_{i|i-1}}{\hat{P}_{i|i-1}~N_n}}~~~~~\forall~~1< i \leq N_s \\
    \end{cases}
\end{equation}

\noindent where $N_n$ is the number of samples in a subset. The overall COV estimate over the required failure probability $(\hat{P}_f)$ is:

\begin{equation}
    \label{eqn:SS_3}
    \hat{\delta} = \sqrt{\sum_{i=1}^{N_s}~\hat{\delta}_i^2}
\end{equation}

\noindent This estimator does not consider correlation between Markov chains and therefore underestimates the COV. A revised estimator that accounts for these correlations has been derived in \citet{Shields2021a}.

\section{Coupled Active Learning, Multifidelity Modeling, and Subset Simulation}\label{sec:AL_MFM}

This section provides an overview of the coupled active learning, multifidelity modeling, and subset simulation algorithm and summarizes it in the form of a flowchart.

\subsection{Adaptive Kriging in Subset Simulation}

Several studies in the past have incorporated Kriging into subset simulation. For example, \citet{Huang2016a} proposed an adaptive Kriging subset simulation (AK-SS) algorithm that works in a similar fashion to that of the AK-MCS algorithm. However, instead of using regular Monte Carlo with the trained Kriging model to estimate failure probabilities, subset simulation was used. \citet{Xu2020a} proposed adaptive Kriging modified subset simulation, which mainly differs from AK-SS in how the intermediate failure thresholds are determined. \citet{Cui2019a} proposed a subset simulation with Kriging and K-means clustering to replace expensive model calls with a cheaper surrogate and identify potential model failure modes, respectively. A commonality between these approaches is that Kriging and subset simulation are de-coupled. That is, although the accuracy of Kriging predictions is checked using a learning function, this is done collectively for an initially generated $N_o$ Monte Carlo samples. In contrast, the approach we employ in this paper does not use such $N_o$ Monte Carlo samples, and instead, Kriging is actively used in subset simulation in place of the HF model with the prediction accuracy checked for every input sample individually.

\begin{figure}[H]
\centering
\includegraphics[scale=0.5]{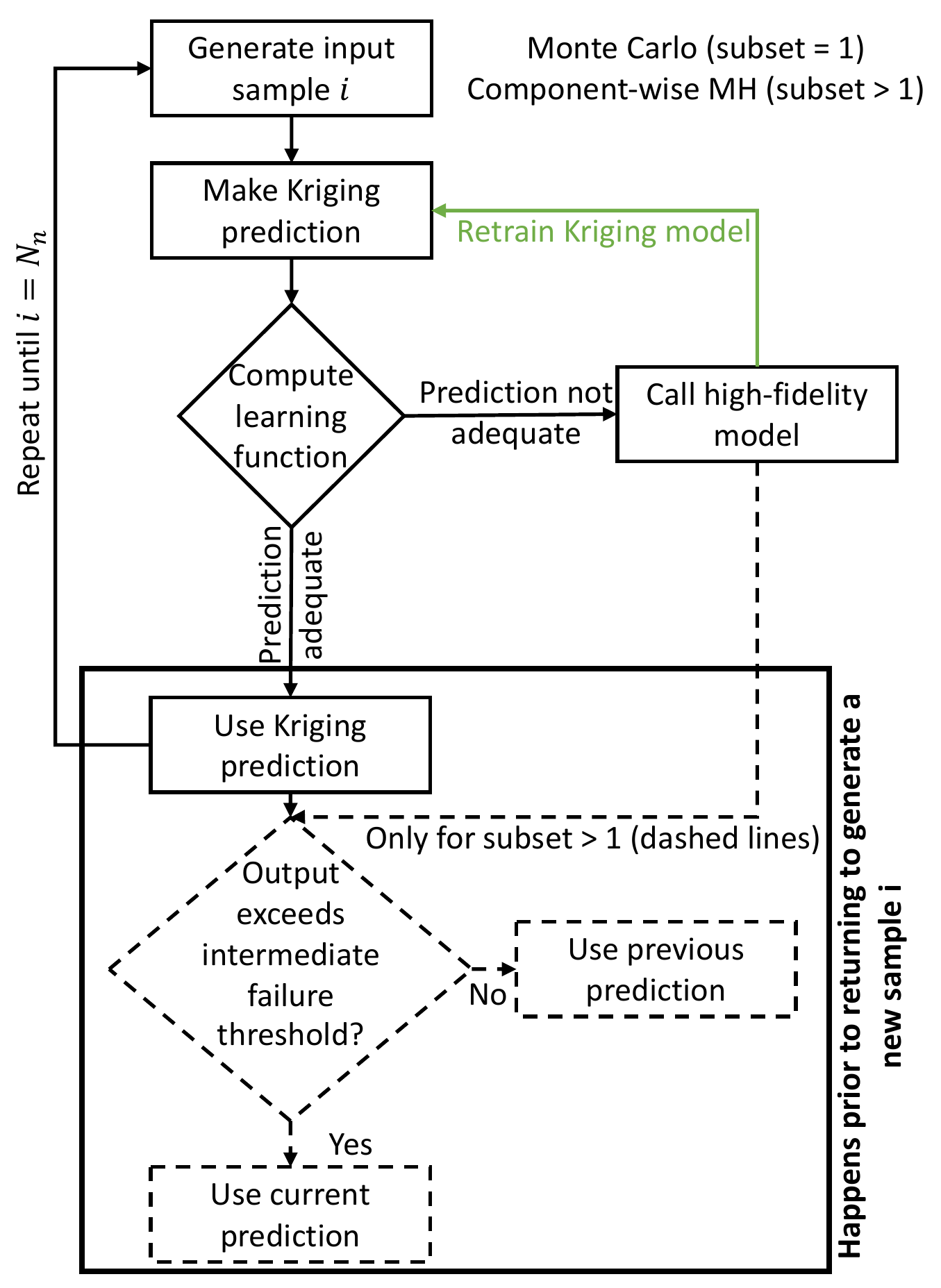}
\caption{Flowchart of the coupled Kriging and subset simulation algorithm.}
\label{SS_GP}
\end{figure}

Figure \ref{SS_GP} presents a flowchart of the coupled Kriging and subset simulation method, which serves as a special case of the multifidelity method proposed in \citet{Dhulipala_AL_MFM} (the full method will be presented in Section 4.3). The method proceeds in a sequential fashion. A Kriging model is first trained on a few evaluations (e.g. about 12) of the HF model. Then, an input sample $\pmb{X}_i$ is generated using a regular Monte Carlo if it is the first subset or a component-wise Metropolis-Hastings (MH) if it is a subset greater than one. An active learning function (e.g., U-function) is computed to evaluate the Kriging performance for input sample $\pmb{X}_i$. If the Kriging prediction is not adequate, the HF model is called and the Kriging model is re-trained by adding input sample $\pmb{X}_i$ to the training set. Otherwise, the Kriging prediction is taken as the model prediction. If it is a subset greater than one, there is an additional step that involves checking whether the output (either from Kriging or HF model) exceeds an intermediate failure threshold $\mathcal{F}_{i-1}$. This additional step is to ensure that the samples follow the conditioning criterion in the distribution $P(\textrm{output} > \mathcal{F}_{i}~|~\textrm{output} > \mathcal{F}_{i-1})$. The procedure is repeated for each subset until $i=N_n$, the required number of samples in a subset.

As can been seen in Figure \ref{SS_GP}, the method is very similar to that of a regular subset simulation, except for the Kriging prediction, performance assessment using a learning function, and re-training steps that are used to replace expensive HF model evaluations with cheaper Kriging predictions. Since Kriging is used in a coupled fashion with a subset simulation, we can also estimate very low probabilities on the order of $10^{-9}$ as demonstrated in Appendix \myref{section_app}. However, as discussed subsequently, special attention should be placed on the learning function as the traditional $U$-function in Equation \eqref{eqn:AKMCS_5} cannot be directly used.

\subsection{Subset-Dependent U Learning Functions}\label{sec:SS_U}

Using the traditional $U$-function, Equation \eqref{eqn:AKMCS_5}, in Figure \ref{SS_GP} can cause the method to break down when the failure probabilities are $10^{-4}$ or less. For smaller failure probabilities, the required failure threshold is usually far away from the nominal distribution of input parameters. Then, the traditional $U$-function, which partly relies on the distance between the Kriging prediction and the failure threshold (the numerator in Equation \eqref{eqn:AKMCS_5}) will be usually larger than $2.0$ and does not trigger calls to the HF model and retraining of the Kriging model. As the subset simulation method proceeds, the under-trained Kriging model will not result in predictions that are close to the required failure threshold as the subset index increases. This causes the method in Figure \ref{SS_GP} to break down and return a zero for the failure probability. 

To combat this breakdown issue, subset-dependent $U$ learning functions were proposed. If the subset index is less than $N_s$ (the final subset), the subset-dependent learning function is defined as:

\begin{equation}
    \label{eqn:U_func1}
    U_i = \frac{|\mu_{{\mathcal{K}}}(\pmb{X})-\mathcal{F}_i|}{\sigma_{{\mathcal{K}}}(\pmb{X})}~~\forall~i<N_s
\end{equation}

\noindent where $\mathcal{F}_i$ is the intermediate failure threshold of subset $i$. The issue with using Equation \eqref{eqn:U_func1} in Figure \ref{SS_GP} is that we do not know $\mathcal{F}_i$ exactly before all the $N_n$ samples in subset $i$ are simulated. A stochastic estimate of $\mathcal{F}_i$ in Equation \eqref{eqn:U_func1} has also been proposed for use. That is, its value is estimated as the $1-p_o$ quantile value of all the sample outputs made until the current sample index in subset $i$ by either the Kriging or HF model. This stochastic estimate of $\mathcal{F}_i$ will quickly converge to the actual value of $\mathcal{F}_i$ with the number of samples in the subset. For the final subset $N_s$, the $U$-function in Equation \eqref{eqn:AKMCS_5} is used.

\subsection{Use of Multifidelity Modeling}\label{sec:MFM}

The coupled Kriging and subset simulation method in Figure \ref{SS_GP} uses a surrogate model to both predict the HF model response and evaluate the prediction quality using active learning functions. Using the concepts of information fusion and adaptation from the multifidelity modeling literature \cite{Peherstorfer2018a}, the method in Figure \ref{SS_GP} can be expanded to decouple the model response prediction and the prediction quality estimation. This allows for physics-based LF models (e.g., using simplified physics or coarse numerical schemes) or data-driven (i.e., computationally cheap surrogate models), or a combination of both to be used for prediction purposes. For our purposes, regardless of whether this is a physics-based model or a data-driven model, we will refer to it as the LF model. A data-driven approach is then used to assess prediction quality, i.e. using Kriging active learning functions. In this multifidelity scheme, we define a corrected prediction from a LF model $(y_{LF}^{C})$ as \cite{Dhulipala_AL_MFM}:

\begin{equation}
    \label{eqn:MFM_1}
    y_{LF}^{C} = y_{LF} + \varepsilon_{*}
\end{equation}

\noindent where, $y_{LF}$ is a LF prediction made by the LF model and $\varepsilon_{*}$ is a machine learned correction term.  The correction model must also quantify the uncertainty associated with the correction in order to leverage active learning functions that decide when $y_{LF}^{C}$ is inadequate and HF model evaluation is necessary. For this reason, we use Kriging to define the correction term $\varepsilon_{*}$ because it is endowed with a natural error measure in its variance estimator. More specifically, $\varepsilon_{*}$ is defined as:

\begin{equation}
    \label{eqn:MFM_2}
    \varepsilon_{*} \approx y_{HF} - y_{LF} \sim p(\varepsilon_{*}~|~\pmb{X},\pmb{X}_{*},\varepsilon)
\end{equation}

\noindent where, $y_{HF}$ is the HF model output, $\pmb{X}_*$ and $\varepsilon = y_{HF}-y_{LF}$ are the training inputs and outputs, respectively, and $\pmb{X}$ are the test inputs. The distribution $p(\varepsilon_{*}~|~\pmb{X},\pmb{X}_{*},\varepsilon)$ in Equation \eqref{eqn:MFM_2} follows a Gaussian with mean $\mu_{\varepsilon_{*}}(\pmb{X})$ and standard deviation $\sigma_{\varepsilon_{*}}(\pmb{X})$. Initially, the Kriging correction model is trained on a few evaluations of both the HF and LF models to characterize $p(\varepsilon_{*}~|~\pmb{X},\pmb{X}_{*},\varepsilon)$. With the use of multifidelity modeling, the subset-dependent $U$ learning functions in Section \ref{sec:SS_U} need to be modified such that the multifidelity subset-dependent $U$ learning function for subset $i<N_s$ is defined as:

\begin{equation}
    \label{eqn:MFM_U_func1}
    U_i = \frac{|y_{LF}+\mu_{{\varepsilon_{*}}}(\pmb{X})-\mathcal{F}_i|}{\sigma_{{\varepsilon_{*}}}(\pmb{X})}~~\forall~i<N_s
\end{equation}

\noindent For $i=N_s$, the final subset, $U$ learning function is defined as:

\begin{equation}
    \label{eqn:MFM_U_func2}
    U_i = \frac{|y_{LF}+\mu_{{\varepsilon_{*}}}(\pmb{X})|}{\sigma_{{\varepsilon_{*}}}(\pmb{X})}~~\forall~i=N_s
\end{equation}

\subsection{Overview of the Method}

\begin{figure}[H]
\centering
\includegraphics[scale=0.5]{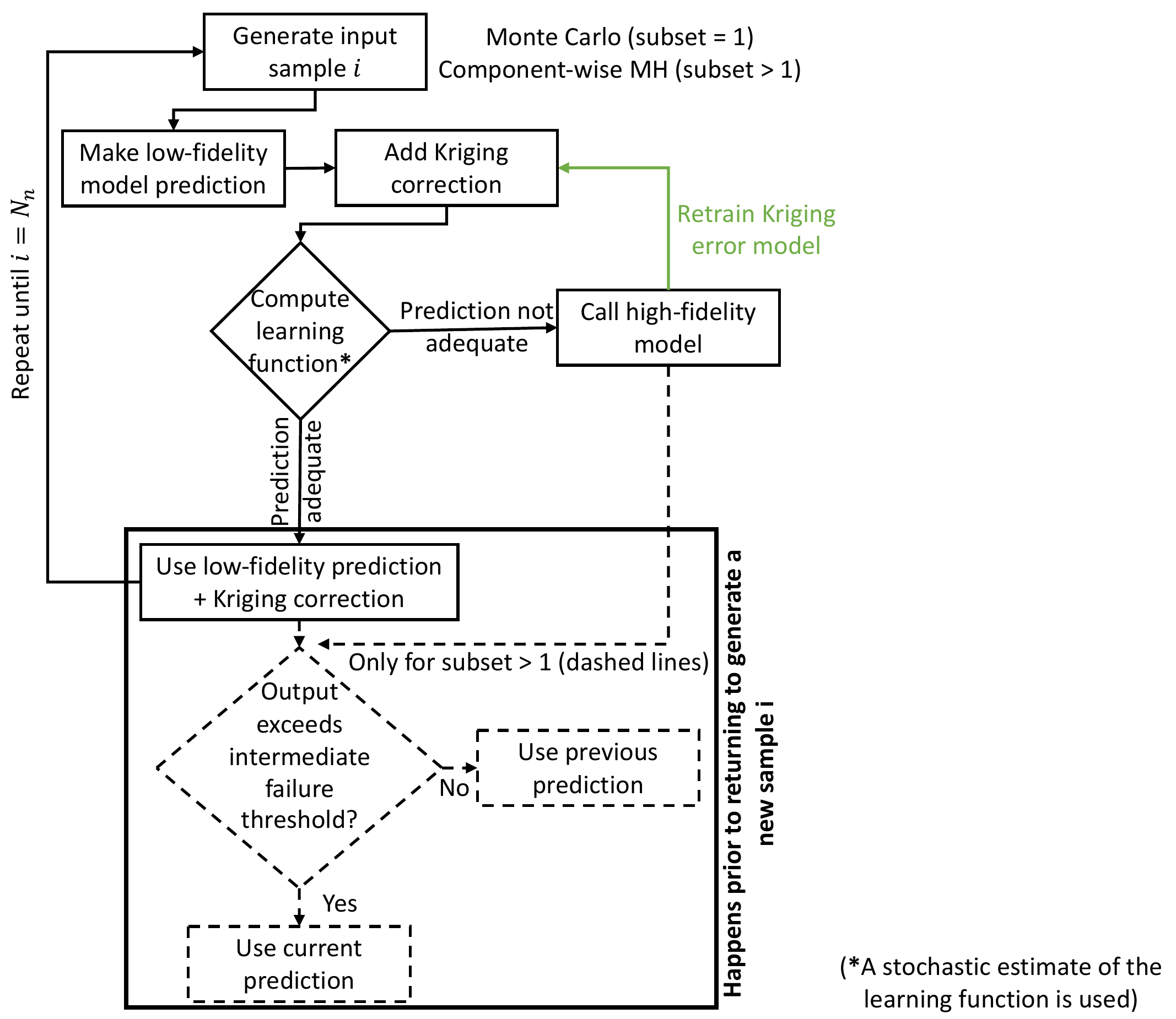}
\caption{Flowchart of the coupled Kriging and subset simulation algorithm with the use of multifidelity modeling.}
\label{SS_GP_MF}
\end{figure}

The method for implementing the coupled Kriging and subset simulation with multifidelity modeling is presented in Figure \ref{SS_GP_MF}. It consists of the following steps:

\begin{enumerate}
    \item Select an LF model. Preferably this will be a physics-based model with either simplified physics or reduced number of degrees of freedom. However, in principal it may also be a data-driven model.
    \item Evaluate both the HF and LF model on a few samples (e.g., 12) and train a Kriging model with multiple length scales to learn the difference between the HF and LF model predictions.
    \item Generate an input parameters sample using a Monte Carlo (for subset $i=1$) or a component-wise MH (for subset $i>1$) and do the following:
    \begin{itemize}
        \item Make an LF prediction.
        \item Compute the Kriging correction using Equation \eqref{eqn:MFM_2} and add it to the LF prediction.
        \item Compute the subset-dependent multifidelity version of the active learning function with Equation \eqref{eqn:MFM_U_func1} or \eqref{eqn:MFM_U_func2} using a stochastic estimate of the failure threshold $i$.
        \item Use the active learning function to decide whether the Kriging-corrected LF prediction is adequate. If not, call the HF model and re-train the Kriging correction model.
        \item If the subset $i>1$, check whether the output exceeds the intermediate failure threshold $\mathcal{F}_{i-1}$. If not, use the previous output value.
    \end{itemize}
    \item Repeat Step 3 until the number of samples in the subset equals $N_s$.
    \item Repeat Steps 3 and 4 for all subsets.
\end{enumerate}

\section{Reliability Estimation of TRISO Nuclear Fuel using 1D Models}\label{sec:1D_TRISO}

In this section, the failure probabilities and reliablities of four 1D TRISO fuel models, which represent the behavior of this fuel under a variety of conditions, will be estimated using the coupled active learning, multifidelity modeling, and subset simulation algorithm previously presented. The 1D model is used here in this initial study as the HF model, but will later be used in Section \ref{sec:2D_TRISO} as the LF model together with a 2D HF model. The TRISO $P_f$ is defined as the product of IPyC layer failure probability $P(\textrm{IPyC})$ and SiC layer failure probability conditioned on IPyC layer failure $P(\textrm{SiC}|\textrm{IPyC})$ as discussed in Section \ref{TRISO_desc}. We use the coupled active learning, multifidelity modeling, and subset simulation algorithm to estimate only $P(\textrm{SiC}|\textrm{IPyC})$ as it is associated with low values of the order $10^{-3}$ to $10^{-5}$. Since $P(\textrm{IPyC})$ is of the order $10^{-1}$, we use the simple AK-MCS algorithm to estimate it. Next, we present the different multifidelity modeling strategies used to estimate $P(\textrm{SiC}|\textrm{IPyC})$ and then discuss the TRISO failure analysis results.

\subsection{Description of the Multifidelity Models}

\begin{figure}[H]
\centering
\includegraphics[width=6.0in, height=3.375in]{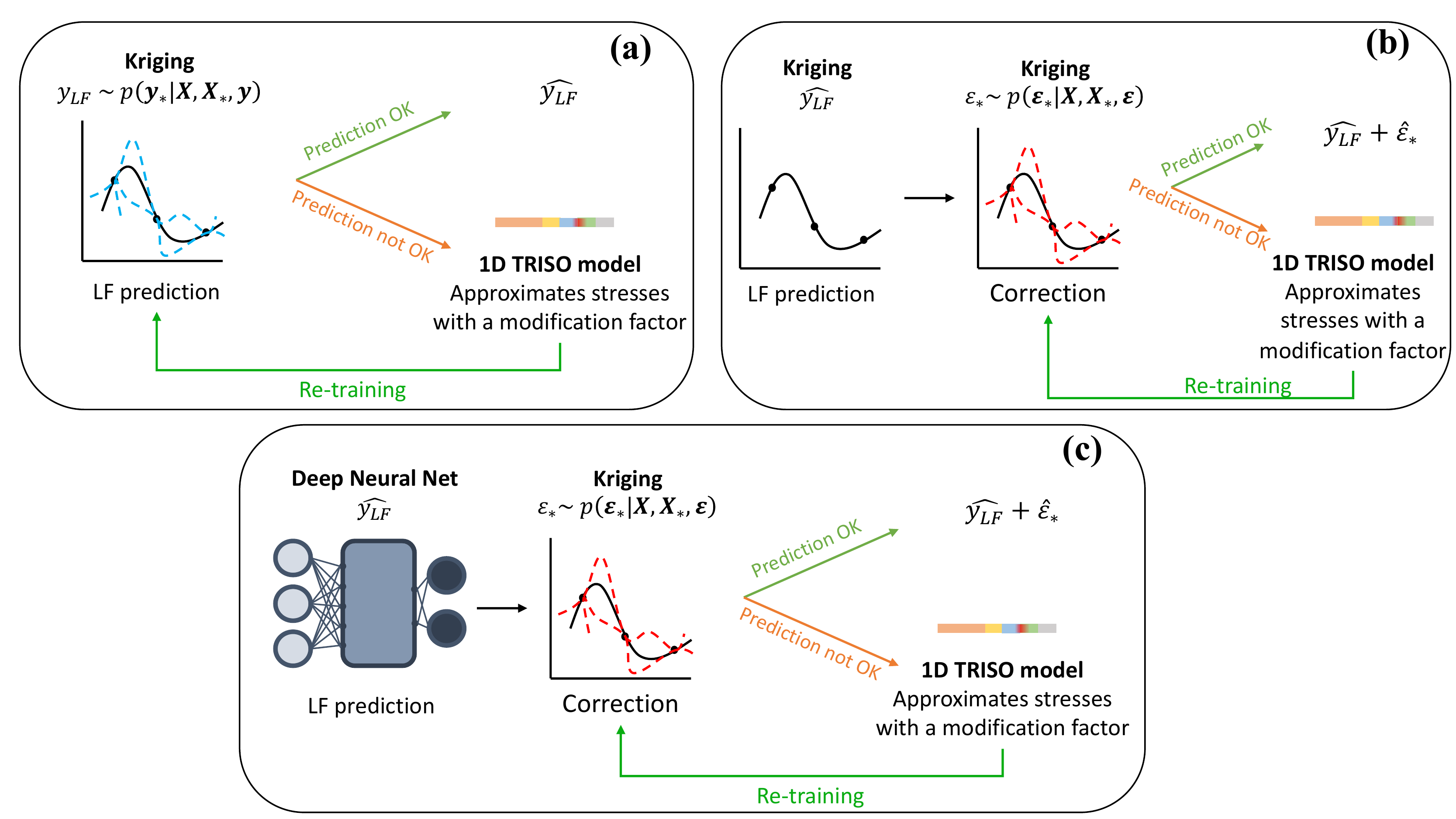}
\caption{Three multifidelity modeling strategies for estimating SiC conditional of IPyC failure probability $[P(\textrm{SiC}|\textrm{IPyC})]$ by approximating the 1D TRISO output: (a) Only Kriging; (b) Kriging LF model output plus an added Kriging correction term; (c) DNN LF model output plus an added Kriging correction term.}
\label{1DTRISO_cases}
\end{figure}

We consider three alternative multifidelity modeling strategies to estimate $P(\textrm{SiC}|\textrm{IPyC})$ for the four TRISO models. Herein, the model output corresponds to the difference between stress and strength in either the IPyC or SiC layer depending upon the failure probability of the layer of interest. In each multifidelity modeling strategy, the LF model is provided as a data-driven model with limited training data. Physics-based LF models will be considered in the next section. Figure \ref{1DTRISO_cases} presents these alternative strategies. Strategy (a) uses the method presented in Sections 4.1 and 4.2 in which a single Kriging model is used to predict the HF (1D TRISO model) outputs and to check prediction accuracy using the active learning functions in Equations \eqref{eqn:U_func1} and \eqref{eqn:AKMCS_5}. Strategy (b) uses a Kriging model with limited training data as the LF model for prediction, and a separate Kriging model for the correction using Equation \eqref{eqn:MFM_1} but also to check the prediction accuracies using the active learning functions in Equations \eqref{eqn:MFM_U_func1} and \eqref{eqn:MFM_U_func2}. Strategy (c) is similar to that of Strategy (b), except that instead of a Kriging model to make an LF prediction, a DNN is used. This DNN has three hidden layers with 20 neurons per layer, and the loss function is regularized using L2 regularization. We used 5000 epochs for training the DNN and a learning rate of 0.002. In Strategy (a), the Kriging model is initially trained on 12 evaluations of the HF 1D TRISO model and is retrained after each HF model evaluation. In Strategies (b) and (c), the LF model is trained on 12 evaluations of the HF 1D TRISO model each and the Kriging correction is retrained after each HF model evaluation.

\subsection{IPyC Layer Failure Results}

The failure probability of the IPyC layer is required to determine the failure probability of the TRISO particle, as discussed in Section \ref{TRISO_desc}. Usually, the IPyC failure probabilities are $O(10^{-1})$. Although in theory, the coupled active learning, multifidelity modeling, and subset simulation algorithm can be used, it is more straightforward to use the AK-MCS algorithm for higher failure probabilities. We used the standard AK-MCS algorithm proposed by \citet{Echard2011a} such that the target COV of $P_f$ is around 0.05. Table \ref{Table_IPyC_failure} presents the results for the four TRISO models discussed in Section \ref{TRISO_unc} along with the reference $P_f$ value from \citet{Dhulipala_TRISO}, who used a Weibull failure theory to compute the failure probabilities. More information about this Weibull approach can be found in either \citet{Dhulipala_TRISO} or \citet{Zok2017a}. Across all four models, not only do the failure probabilities and reliability indices compare well between the AK-MCS and reference values, but also, the AK-MCS only requires a small fraction of calls to the HF 1D TRISO model compared to the total number of input samples for a target COV of 0.05.

\begin{table}[H]
\centering
\caption{1D TRISO model IPyC layer failure probabilities computed using the standard AK-MCS algorithm \cite{Echard2011a}.}
\label{Table_IPyC_failure}
\small
\begin{tabular}{ |c|c|c|c|c| }
\hline
\multicolumn{5}{|c|}{\textbf{IPyC layer failure probability for the 1D TRISO models}}\\
\hline
& \Centerstack[c]{\textbf{Model 1}} & \Centerstack[c]{\textbf{Model 2}} & \Centerstack[c]{\textbf{Model 3}} & \Centerstack[c]{\textbf{Model 4}}\\
\hline
\Centerstack[c]{\textbf{AK-MCS} \\ \textbf{method \cite{Echard2011a}}} & \Centerstack[c]{$\hat{P}_f=0.226$ \\ $\hat{\delta}_{P_f}=0.047$ \\ $\hat{\beta} = 0.752$ \\ \# HF calls $=43$ \\ \# samples $=1500$} & \Centerstack[c]{$\hat{P}_f=0.115$ \\ $\hat{\delta}_{P_f}=0.047$ \\ $\hat{\beta} = 1.2$ \\ \# HF calls $=56$ \\ \# samples $=3500$} & \Centerstack[c]{$\hat{P}_f=0.98$ \\ $\hat{\delta}_{P_f}=0.015$ \\ $\hat{\beta} =$ N/A \\ \# HF calls $=12$ \\ \# samples $=90$} & \Centerstack[c]{$\hat{P}_f=0.082$ \\ $\hat{\delta}_{P_f}=0.045$ \\ $\hat{\beta} = 1.391$ \\ \# HF calls $=84$ \\ \# samples $=5500$}\\
\hline
\Centerstack[c]{\textbf{Reference result} \\ \textbf{using the Weibull} \\ \textbf{approach from \cite{Dhulipala_TRISO}}} & \Centerstack[c]{$\hat{P}_f=0.228$ \\ $\hat{\delta}_{P_f}=0.025$ \\ $\hat{\beta} = 0.745$} & \Centerstack[c]{$\hat{P}_f=0.104$ \\ $\hat{\delta}_{P_f}=0.026$ \\ $\hat{\beta} = 1.26$} & \Centerstack[c]{$\hat{P}_f=0.967$ \\ $\hat{\delta}_{P_f}=0.004$ \\ $\hat{\beta} = $ N/A} & \Centerstack[c]{$\hat{P}_f=0.086$ \\ $\hat{\delta}_{P_f}=0.035$ \\ $\hat{\beta} = 1.36$}\\
\hline
\end{tabular}
\centering
\begin{tablenotes}
\item[] \textbf{Notations.} $\hat{P}_f$: Failure probability estimate of the IPyC layer; $\hat{\delta}_{{P}_f}$: COV estimate over $\hat{P}_f$; $\hat{\beta}$: Reliability index estimate
\end{tablenotes}
\end{table}

\subsection{SiC Layer Failure Results}

The TRISO particle $P_f$ is defined in Equation \eqref{eqn:TRISO_Pf}, wherein the $P(SiC|IPyC)$ and $P(IPyC)$ are computed separately and then multiplied to get the final failure probability. First, $P(SiC|IPyC)$ results using the active learning, mulitfidelity modeling, and subset simulation method in Figure \ref{SS_GP_MF} are discussed for the four TRISO models. Then, the overall TRISO failure probability results are discussed.

\subsubsection{SiC Failure Conditioned on IPyC Failure}

SiC failures conditioned on IPyC failure probabilities for the four TRISO models are computed using the method presented in Figure \ref{SS_GP_MF}. Per Figure \ref{1DTRISO_cases}, three multifidelity modeling strategies are used: (a) only Kriging; (b) Kriging LF prediction with Kriging correction and (c) DNN LF prediction with Kriging correction. We used 5,000 samples in each subset such that the target COV over $P_f$ is about $0.075$. Given that the IPyC failure probabilities have a COV of around $0.05$, the resultant COV over the TRISO failure probabilities will be around $0.1$. A COV value of $0.1$ is considered satisfactory for the TRISO model case study \cite{Dhulipala_TRISO,Jiang2021a,Miller2018a}. While for Models 1, 2, and 4, we used four subsets, for Model 3, we used three. Table \ref{Table_SiC_IPyC_failure} presents the results using these three multifidelity modeling strategies along with the reference results from \citet{Dhulipala_TRISO}. The number of HF calls in Table \ref{Table_SiC_IPyC_failure} include the calls made during the initial training of all the LF models. Overall, all the three multifidelity modeling strategies are satisfactorily predicting the SiC failure conditioned on IPyC failure probabilities and the corresponding reliability indices given the COV value. However, some interesting observations can be made on the number of calls to the 1D TRISO model. The use of either Kriging (i.e., Strategy (b)) or DNN (i.e., Strategy (c)) to make an LF prediction and then Kriging to make a correction always results in fewer calls to the 1D TRISO model than using Kriging alone (i.e., Strategy (a)). Strategy (c), which uses a DNN, requires the fewest calls to the 1D TRISO model, with the total calls being {$26\%$ and $18\%$} less for all four TRISO models for Strategies (a) and (b), respectively. These observations imply that: (1) using a single LF model with a ML correction can more effectively reduce the computational expense than using a single surrogate model; and (2) the approximation quality of the LF model when using a corrected LF model can impact the computational gains.

\begin{table}[H]
\centering
\caption{1D TRISO model SiC conditional on IPyC failure probabilities computed using the coupled active learning, multifidelity modeling, and subset simulation algorithm. Results corresponding to the three multifidelity modeling strategies in Figure \ref{1DTRISO_cases} are presented.}
\label{Table_SiC_IPyC_failure}
\small
\begin{tabular}{ |c|c|c|c|c| }
\hline
\multicolumn{5}{|c|}{\textbf{SiC layer failure probability conditional on IPyC layer failure for the 1D TRISO models}}\\
\hline
& \Centerstack[c]{\textbf{Model 1}} & \Centerstack[c]{\textbf{Model 2}} & \Centerstack[c]{\textbf{Model 3}} & \Centerstack[c]{\textbf{Model 4}}\\
\hline
\Centerstack[c]{\textbf{Multifidelity} \\ \textbf{strategy (a)} \\ \textbf{in Figure \ref{1DTRISO_cases}:} \\ \\ Only Kriging} & \Centerstack[c]{$\hat{P}_f=5.37E-4$ \\ $\hat{\delta}_{P_f}=0.075$ \\ $\hat{\beta} = 3.27$ \\ \# HF calls $=296$ \\ \# samples $=20,000$} & \Centerstack[c]{$\hat{P}_f=3.4E-4$ \\ $\hat{\delta}_{P_f}=0.076$ \\ $\hat{\beta} = 3.39$ \\ \# HF calls $=274$ \\ \# samples $=20,000$} & \Centerstack[c]{$\hat{P}_f=1.1E-3$ \\ $\hat{\delta}_{P_f}=0.072$ \\ $\hat{\beta} = 3.06$ \\ \# HF calls $=331$ \\ \# samples $=15,000$} & \Centerstack[c]{$\hat{P}_f=1.7E-4$ \\ $\hat{\delta}_{P_f}=0.079$ \\ $\hat{\beta} = 3.58$ \\ \# HF calls $=541$ \\ \# samples $=20,000$}\\
\hline
\Centerstack[c]{\textbf{Multifidelity} \\ \textbf{strategy (b)} \\ \textbf{in Figure \ref{1DTRISO_cases}:} \\ \\ Kriging + Kriging} & \Centerstack[c]{$\hat{P}_f=4.63E-4$ \\ $\hat{\delta}_{P_f}=0.075$ \\ $\hat{\beta} = 3.31$ \\ \# HF calls $=281$ \\ \# samples $=20,000$} & \Centerstack[c]{$\hat{P}_f=2.98E-4$ \\ $\hat{\delta}_{P_f}=0.076$ \\ $\hat{\beta} = 3.43$ \\ \# HF calls $=229$ \\ \# samples $=20,000$} & \Centerstack[c]{$\hat{P}_f=1.1E-3$ \\ $\hat{\delta}_{P_f}=0.072$ \\ $\hat{\beta} = 3.06$ \\ \# HF calls $=301$ \\ \# samples $=15,000$} & \Centerstack[c]{$\hat{P}_f=1.6E-4$ \\ $\hat{\delta}_{P_f}=0.08$ \\ $\hat{\beta} = 3.59$ \\ \# HF calls $=504$ \\ \# samples $=20,000$}\\
\hline
\Centerstack[c]{\textbf{Multifidelity} \\ \textbf{strategy (c)} \\ \textbf{in Figure \ref{1DTRISO_cases}:} \\ \\ DNN + Kriging} & \Centerstack[c]{$\hat{P}_f=4.18E-4$ \\ $\hat{\delta}_{P_f}=0.075$ \\ $\hat{\beta} = 3.34$ \\ \# HF calls $=220$ \\ \# samples $=20,000$} & \Centerstack[c]{$\hat{P}_f=3.47E-4$ \\ $\hat{\delta}_{P_f}=0.076$ \\ $\hat{\beta} = 3.39$ \\ \# HF calls $=197$ \\ \# samples $=20,000$} & \Centerstack[c]{$\hat{P}_f=0.93E-3$ \\ $\hat{\delta}_{P_f}=0.074$ \\ $\hat{\beta} = 3.11$ \\ \# HF calls $=261$ \\ \# samples $=15,000$} & \Centerstack[c]{$\hat{P}_f=2.06E-4$ \\ $\hat{\delta}_{P_f}=0.079$ \\ $\hat{\beta} = 3.53$ \\ \# HF calls $=396$ \\ \# samples $=20,000$}\\
\hline
\Centerstack[c]{\textbf{Reference result} \\ \textbf{using the Weibull} \\ \textbf{approach from \cite{Dhulipala_TRISO}}} & \Centerstack[c]{$\hat{P}_f=4.65E-4$ \\ $\hat{\delta}_{P_f}=0.039$ \\ $\hat{\beta} = 3.31$} & \Centerstack[c]{$\hat{P}_f=3.27E-4$ \\ $\hat{\delta}_{P_f}=0.039$ \\ $\hat{\beta} = 3.41$} & \Centerstack[c]{$\hat{P}_f=0.86E-3$ \\ $\hat{\delta}_{P_f}=0.063$ \\ $\hat{\beta} = 3.13$} & \Centerstack[c]{$\hat{P}_f=1.52E-4$ \\ $\hat{\delta}_{P_f}=0.063$ \\ $\hat{\beta} = 3.61$}\\
\hline
\end{tabular}
\centering
\begin{tablenotes}
\item[] \textbf{Notations.} $\hat{P}_f$: Failure probability estimate of SiC layer conditional on IPyC layer failure; $\hat{\delta}_{{P}_f}$: COV estimate over $\hat{P}_f$; $\hat{\beta}$: Reliability index estimate
\item[] \textbf{Abbreviation.} DNN: Deep Neural Network
\end{tablenotes}
\end{table}

\begin{figure}[H]
\begin{subfigure}{0.5\textwidth}
\centering
\includegraphics[width=2.75in, height=2.5in]{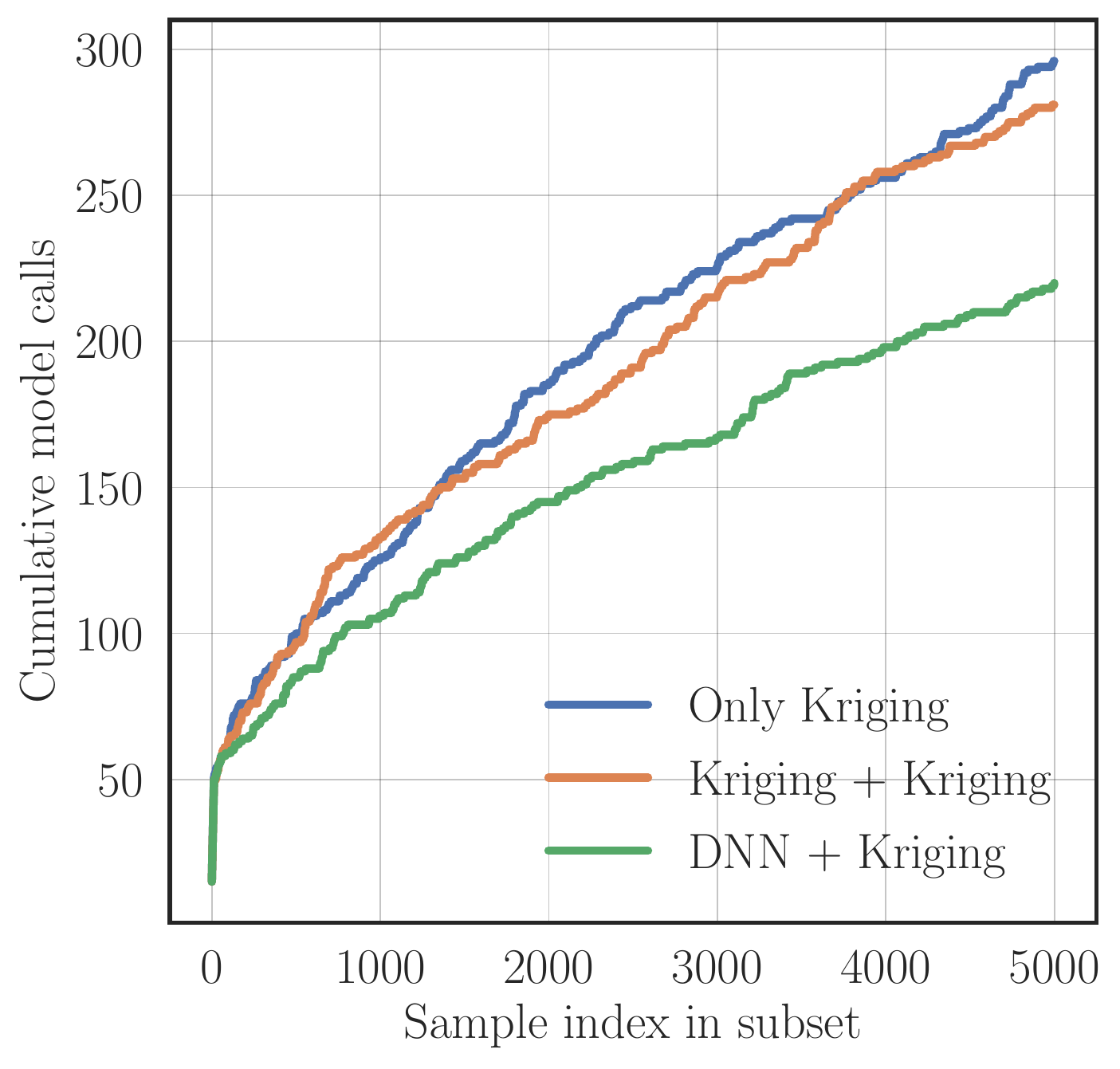}
\caption{}
\label{1DTRISO_calls_1}
\end{subfigure}
\begin{subfigure}{0.5\textwidth}
\centering
\includegraphics[width=2.75in, height=2.5in]{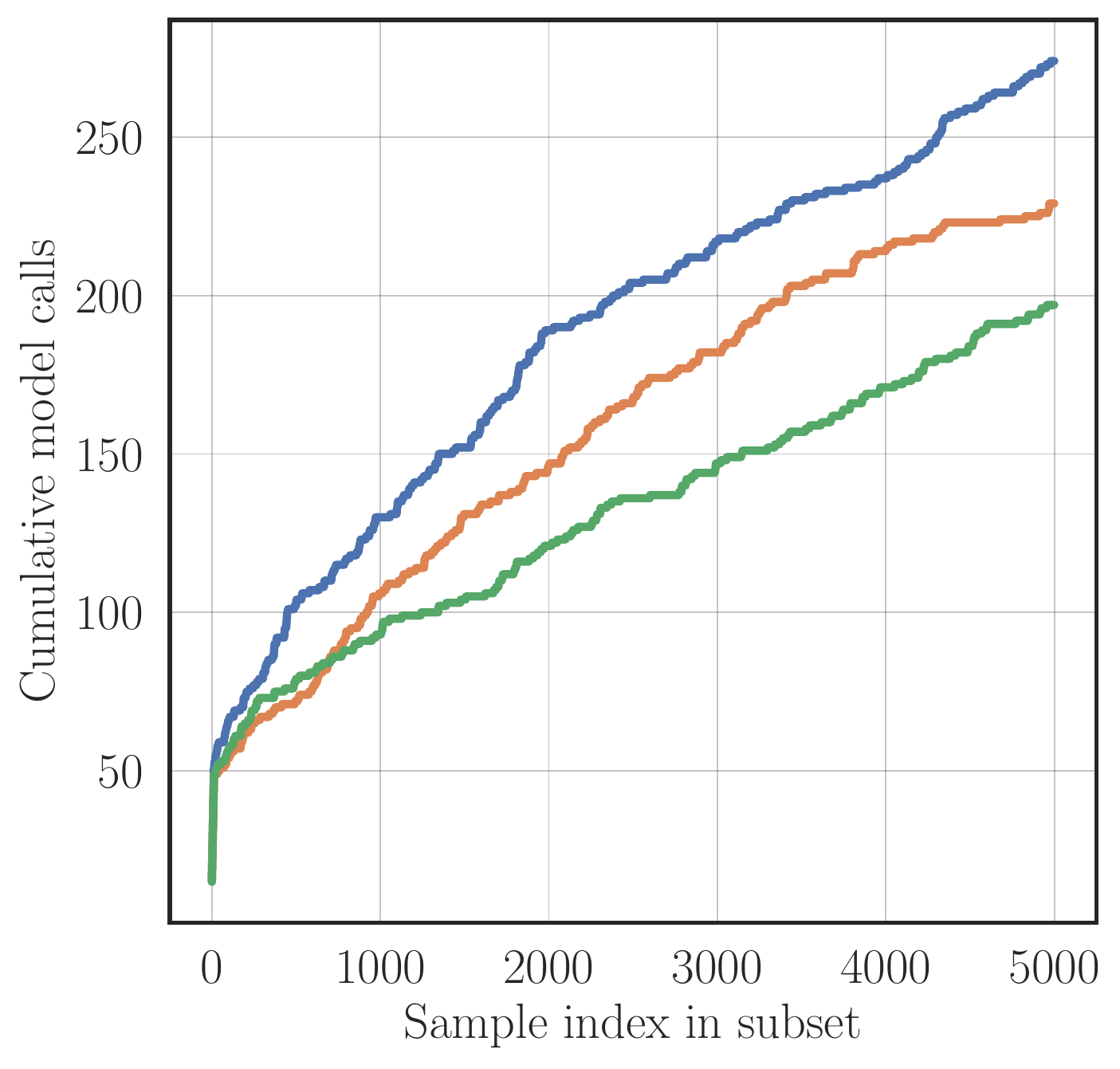}
\caption{}
\label{1DTRISO_calls_2}
\end{subfigure}
\begin{subfigure}{0.5\textwidth}
\centering
\includegraphics[width=2.75in, height=2.5in]{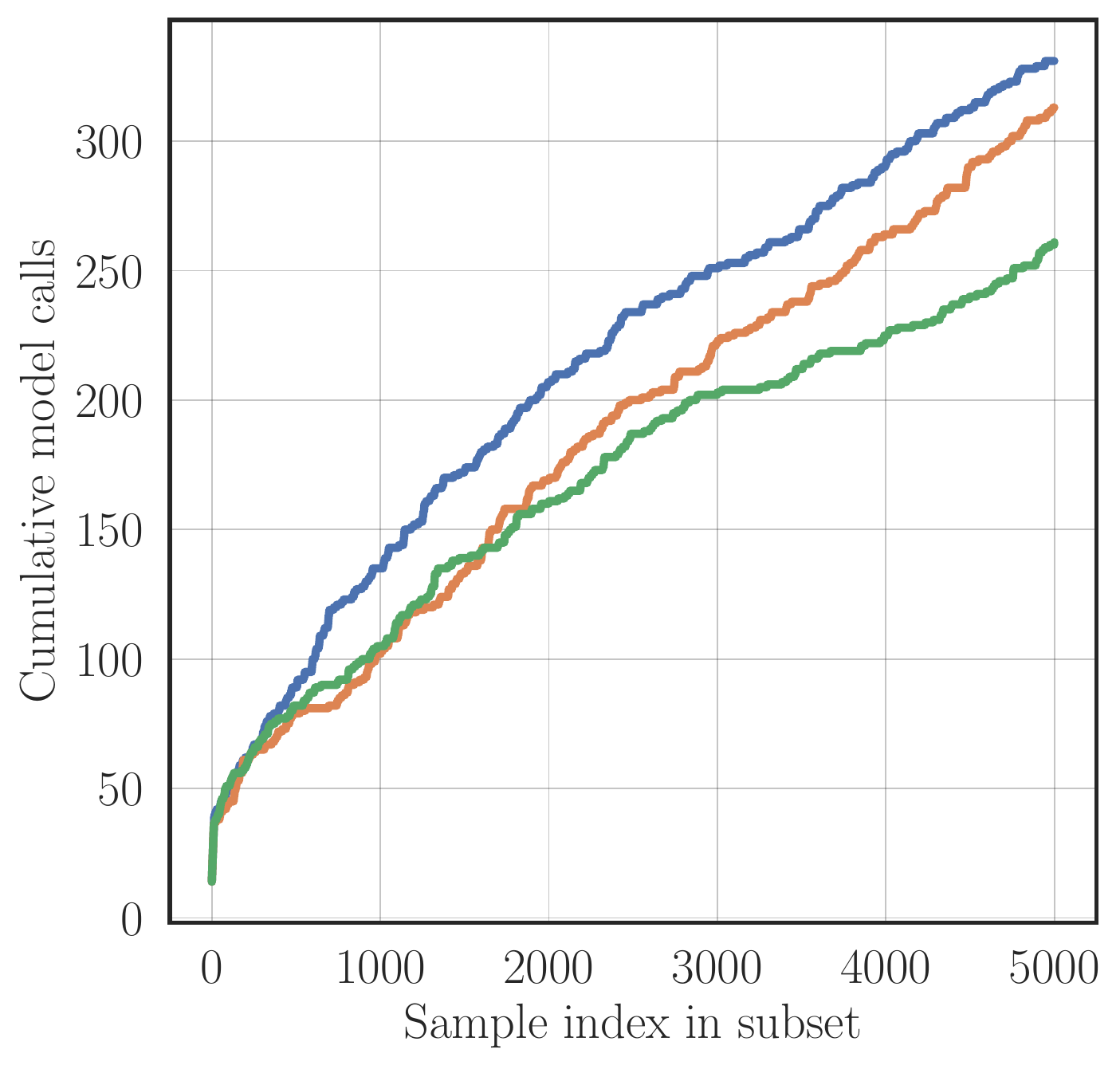}
\caption{}
\label{1DTRISO_calls_3}
\end{subfigure}
\begin{subfigure}{0.5\textwidth}
\centering
\includegraphics[width=2.75in, height=2.5in]{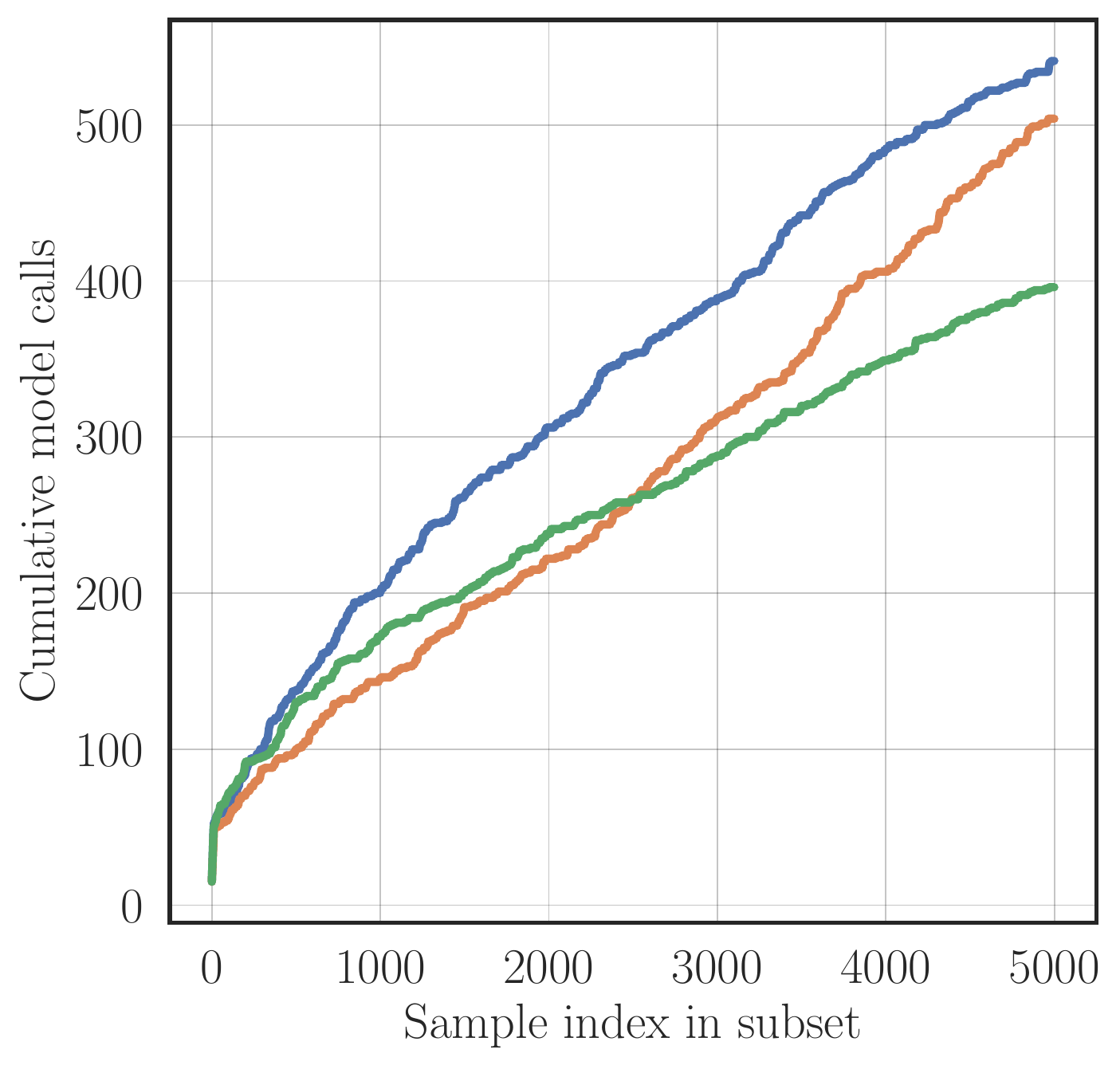}
\caption{}
\label{1DTRISO_calls_4}
\end{subfigure}
\caption{Evolution of the cumulative number of HF model (1D TRISO model) calls with the sample index in the subset for the three multifidelity modeling strategies in Figure \ref{1DTRISO_cases}. (a) Subset 1, (b) Subset 2, (c) Subset 3, (d) Subset 4. Since there are four subsets, at each sample index, the number of HF model calls across these subsets are added.}
\label{1DTRISO_calls}
\end{figure}

Figure \ref{1DTRISO_calls} plots the cumulative number of 1D TRISO model calls as a function of the sample index in the subset for the four TRISO models for each of the multifidelity strategies employed here. Since there are several subsets, the number of model calls across all the subsets are added at each sample index. A possible reason why the strategies with multiple LF models are consistently calling the HF model less often is due to a greater information gain. In Strategies (b) and (c) in Figure \ref{1DTRISO_cases}, the LF model provides an approximation to the HF model output which the Kriging correction model only has to correct. Both the LF models are contributing information regarding the HF output. However, in Strategy (a) in Figure \ref{1DTRISO_cases}, only one LF model (i.e., the Kriging model) is contributing information regarding the HF output. Between Strategies (b) and (c), the fact that the the use of DNN prediction and Kriging correction consistently requires fewer calls to the HF model may be due to better generalization. That is, the DNN model (Strategy [c]) has been trained with a regularization term in the loss function which helps mitigate over-fitting. The Kriging model (Strategy [b]) over-fits the training data set and hence results in more calls to the HF model as the algorithm proceeds. One way to somewhat mitigate this over-fitting issue is to add a noise variance term to the LF Kriging predictions in Equations \eqref{eqn:Kriging_1}--\eqref{eqn:Kriging_4} with an assumed noise value.

\subsubsection{Overall SiC (and TRISO Fuel) Failure}

Following Equation \eqref{eqn:TRISO_Pf}, the IPyC failure and SiC failure conditioned on IPyC failure probabilities, respectively, from Tables \ref{Table_IPyC_failure} and \ref{Table_SiC_IPyC_failure} are multiplied. Table \ref{Table_SiC_failure} presents the overall TRISO failure probability results for the four TRISO models. Overall, the results from the three multifidelity modeling strategies compare satisfactorily with the reference results, given that the resultant COV is around 0.1. In line with the previous observations, Strategy (c) with DNN prediction and Kriging correction requires the fewest HF model calls across all the TRISO models.

\begin{table}[H]
\centering
\caption{1D TRISO model overall SiC (and TRISO fuel) failure probabilities computed by combining the results from Tables \ref{Table_IPyC_failure} and \ref{Table_SiC_IPyC_failure} per Equation \eqref{eqn:TRISO_Pf}. Results corresponding to the three multifidelity modeling strategies in Figure \ref{1DTRISO_cases} are presented.}
\label{Table_SiC_failure}
\small
\begin{tabular}{ |c|c|c|c|c| }
\hline
\multicolumn{5}{|c|}{\textbf{Overall SiC layer (and TRISO) failure probability for the 1D TRISO models}}\\
\hline
& \Centerstack[c]{\textbf{Model 1}} & \Centerstack[c]{\textbf{Model 2}} & \Centerstack[c]{\textbf{Model 3}} & \Centerstack[c]{\textbf{Model 4}}\\
\hline
\Centerstack[c]{\textbf{AK-MCS +} \\ \textbf{Multifidelity} \\ \textbf{Strategy (a)} \\ \textbf{in Figure \ref{1DTRISO_cases}}} & \Centerstack[c]{$\hat{P}_f=1.21E-4$ \\ $\hat{\delta}_{P_f}=0.089$ \\ $\hat{\beta} = 3.67$ \\ \# HF calls $=339$ \\ \# samples $=21,500$} & \Centerstack[c]{$\hat{P}_f=3.91E-5$ \\ $\hat{\delta}_{P_f}=0.089$ \\ $\hat{\beta} = 3.94$ \\ \# HF calls $=330$ \\ \# samples $=23,500$} & \Centerstack[c]{$\hat{P}_f=1.08E-3$ \\ $\hat{\delta}_{P_f}=0.074$ \\ $\hat{\beta} = 3.07$ \\ \# HF calls $=343$ \\ \# samples $=15,090$} & \Centerstack[c]{$\hat{P}_f=1.39E-5$ \\ $\hat{\delta}_{P_f}=0.091$ \\ $\hat{\beta} = 4.19$ \\ \# HF calls $=625$ \\ \# samples $=25,500$}\\
\hline
\Centerstack[c]{\textbf{AK-MCS +} \\ \textbf{Multifidelity} \\ \textbf{Strategy (b)} \\ \textbf{in Figure \ref{1DTRISO_cases}}} & \Centerstack[c]{$\hat{P}_f=1.04E-4$ \\ $\hat{\delta}_{P_f}=0.089$ \\ $\hat{\beta} = 3.71$ \\ \# HF calls $=324$ \\ \# samples $=21,500$} & \Centerstack[c]{$\hat{P}_f=3.43E-5$ \\ $\hat{\delta}_{P_f}=0.089$ \\ $\hat{\beta} = 3.98$ \\ \# HF calls $=285$ \\ \# samples $=23,500$} & \Centerstack[c]{$\hat{P}_f=1.08E-3$ \\ $\hat{\delta}_{P_f}=0.074$ \\ $\hat{\beta} = 3.07$ \\ \# HF calls $=313$ \\ \# samples $=15,090$} & \Centerstack[c]{$\hat{P}_f=1.31E-5$ \\ $\hat{\delta}_{P_f}=0.092$ \\ $\hat{\beta} = 4.2$ \\ \# HF calls $=588$ \\ \# samples $=25,500$}\\
\hline
\Centerstack[c]{\textbf{AK-MCS +} \\ \textbf{Multifidelity} \\ \textbf{Strategy (c)} \\ \textbf{in Figure \ref{1DTRISO_cases}}} & \Centerstack[c]{$\hat{P}_f=0.94E-4$ \\ $\hat{\delta}_{P_f}=0.089$ \\ $\hat{\beta} = 3.73$ \\ \# HF calls $=263$ \\ \# samples $=21,500$} & \Centerstack[c]{$\hat{P}_f=3.99E-5$ \\ $\hat{\delta}_{P_f}=0.089$ \\ $\hat{\beta} = 3.94$ \\ \# HF calls $=253$ \\ \# samples $=23,500$} & \Centerstack[c]{$\hat{P}_f=0.91E-3$ \\ $\hat{\delta}_{P_f}=0.076$ \\ $\hat{\beta} = 3.12$ \\ \# HF calls $=273$ \\ \# samples $=15,090$} & \Centerstack[c]{$\hat{P}_f=1.77E-5$ \\ $\hat{\delta}_{P_f}=0.091$ \\ $\hat{\beta} = 4.13$ \\ \# HF calls $=480$ \\ \# samples $=25,500$}\\
\hline
\Centerstack[c]{\textbf{Reference result} \\ \textbf{using the Weibull} \\ \textbf{approach from \cite{Dhulipala_TRISO}}} & \Centerstack[c]{$\hat{P}_f=1.06E-4$ \\ $\hat{\delta}_{P_f}=0.046$ \\ $\hat{\beta} = 3.7$} & \Centerstack[c]{$\hat{P}_f=3.4E-5$ \\ $\hat{\delta}_{P_f}=0.047$ \\ $\hat{\beta} = 3.98$} & \Centerstack[c]{$\hat{P}_f=0.83E-3$ \\ $\hat{\delta}_{P_f}=0.063$ \\ $\hat{\beta} = 3.15$} & \Centerstack[c]{$\hat{P}_f=1.3E-5$ \\ $\hat{\delta}_{P_f}=0.072$ \\ $\hat{\beta} = 4.21$}\\
\hline
\end{tabular}
\centering
\begin{tablenotes}
\item[] \textbf{Notations.} $\hat{P}_f$: Failure probability estimate of TRISO; $\hat{\delta}_{{P}_f}$: COV estimate over $\hat{P}_f$; $\hat{\beta}$: Reliability index estimate
\end{tablenotes}
\end{table}

\section{Reliability Estimation of TRISO Nuclear Fuel using 2D and 1D Models}\label{sec:2D_TRISO}

Instead of relying on only a 1D TRISO model to approximate the failure probability, in this section, we consider a 2D TRISO model as a HF model and the 1D model as the LF model. We pick Models 1 and 2 from the discussion in Section \ref{TRISO_unc}. Moreover, the emphasis of this section is to further evaluate the performance of the coupled active learning, multifidelity modeling, and subset simulation method on a more expensive model and not to compare the failure probability results between 1D and 2D TRISO models. The results of the 2D TRISO model can be sensitive to the mesh density of the finite element model. The stress concentration in the SiC layer due to the crack in the IPyC layer, which is represented using XFEM, is sensitive to the mesh resolution near the crack region. Additionally, the accuracy of the 1D TRISO failure probability results depend on the accuracy of the stress modification factors that are used to transform the 1D model stresses to 2D stresses accounting for IPyC cracking. \citet{Dhulipala_TRISO} and \citet{Jiang2021b} conduct a detailed investigation on the similarities and differences between 1D and 2D TRISO results. In this section, the model used to predict SiC failure due to stress concentration caused by IPyC cracking has a mesh density that results in each model evaluation taking about 4 minutes on ten processors. We will next discuss the strategies for multifidelity modeling, IPyC failure results, and overall SiC (and TRISO) failure results. We will conclude this section with a discussion on the computational budget.

\subsection{Description of the Multifidelity Models}

\begin{figure}[H]
\centering
\includegraphics[width=4.0in, height=3.375in]{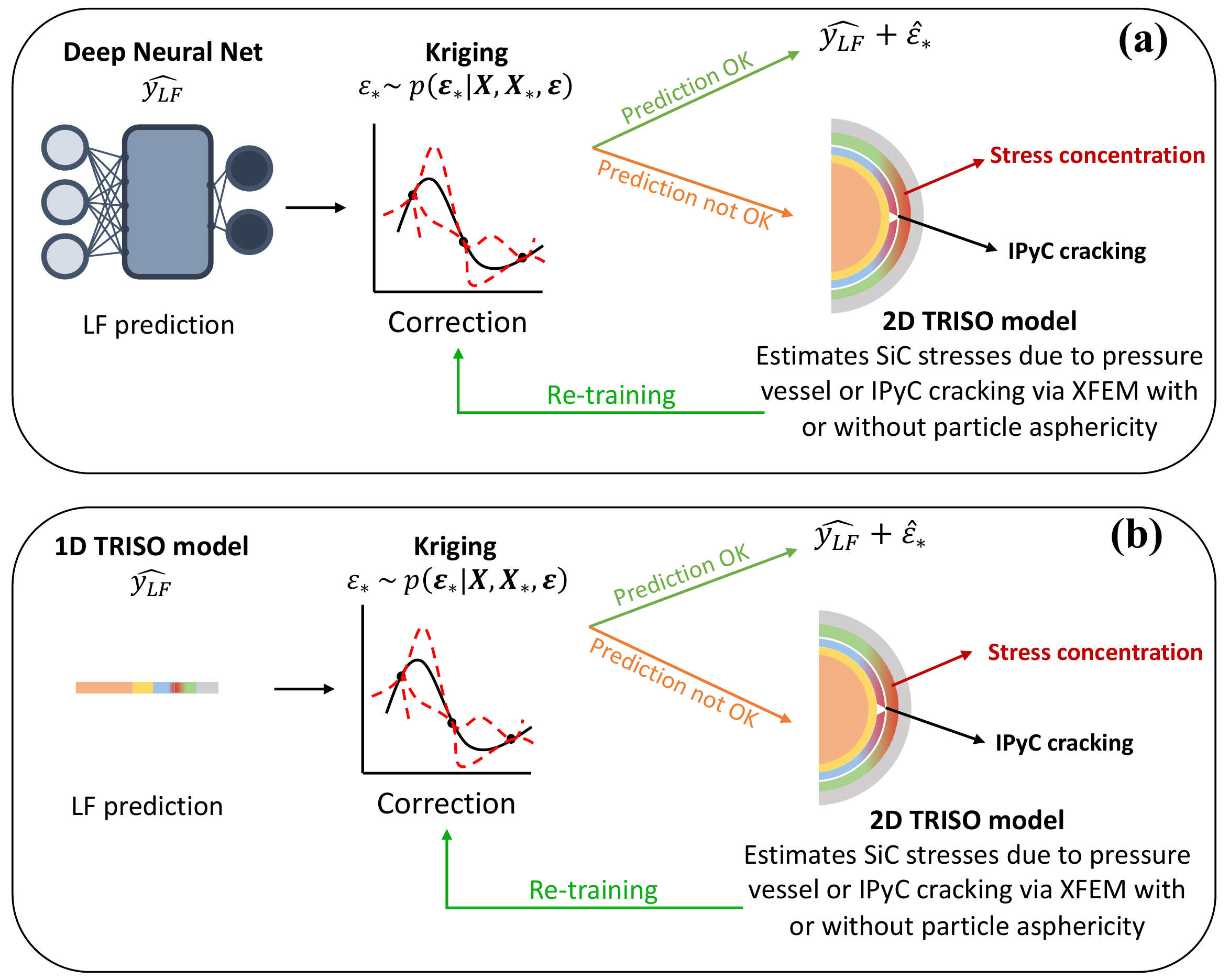}
\caption{Two multifidelity modeling strategies for estimating SiC conditional of IPyC failure probability $[P(\textrm{SiC}|\textrm{IPyC})]$ by approximating the 2D TRISO output: (a) DNN LF model output plus an added Kriging correction term; and (b) 1D TRISO LF model output plus an added Kriging correction term. While (a) is a data-driven strategy, (b) is a physics-based strategy.}
\label{2DTRISO_cases}
\end{figure}

For modeling SiC failure conditioned on IPyC failure with the 2D model, we use two multifidelity modeling strategies. From the results corresponding to the 1D TRISO model in Section \ref{sec:1D_TRISO}, using a DNN led to not only accurate failure probability estimates but also the fewest calls to the HF model. Therefore, the first strategy, as presented in Figure \ref{2DTRISO_cases}(a), consists of a DNN trained on 12 evaluations of the 2D TRISO model, which makes the LF prediction. This DNN is composed of three hidden layers with 20 neurons per layer, and its loss function is regularized using L2 regularization. We used 5000 epochs for training the DNN and a learning rate of 0.002. Then, a Kriging correction term is added to the LF prediction from the DNN. The second strategy, as presented in Figure \ref{2DTRISO_cases}(b), consists of a 1D TRISO model that makes the LF prediction with an added Kriging correction term. Overall, while Strategy (a) is ``data-driven'' because of the use of a DNN, Strategy (b) is ``physics-based'' because of the use of a 1D TRISO model. We then compare the performance of these approaches in terms of the number of times the 2D TRISO model is called.

\subsection{IPyC Layer and Overall SiC Layer (and TRISO) Failure Results}

In line with the approach adopted in Section \ref{sec:1D_TRISO}, the IPyC failure probability is computed using the standard AK-MCS method \cite{Echard2011a} because its value is $O(10^{-1})$. Table \ref{Table_2D_IPyC_failure} presents the results. Both the failure probability and reliability index estimates match well across the two models with the reference results using a Weibull approach from \citet{Dhulipala_TRISO}, taking into consideration the COV over all the results. In addition, the AK-MCS requires a small fraction of calls to the 2D TRISO model in comparison with the number of input samples evaluated.

\begin{table}[H]
\centering
\caption{2D TRISO model IPyC layer failure probability computed using the standard AK-MCS algorithm \cite{Echard2011a}.}
\label{Table_2D_IPyC_failure}
\small
\begin{tabular}{ |c|c|c| }
\hline
\multicolumn{3}{|c|}{\textbf{IPyC layer failure probability for the 2D TRISO models}}\\
\hline
& \textbf{Model 1} & \textbf{Model 2} \\
\hline
\Centerstack[c]{\textbf{AK-MCS} \\ \textbf{method \cite{Echard2011a}}} & \Centerstack[c]{$\hat{P}_f=0.278$ \\ $\hat{\delta}_{P_f}=0.042$ \\ $\hat{\beta} = 0.587$ \\ \# HF calls $=36$ \\ \# samples $=1500$} & \Centerstack[c]{$\hat{P}_f=0.125$ \\ $\hat{\delta}_{P_f}=0.044$ \\ $\hat{\beta} = 1.15$ \\ \# HF calls $=52$ \\ \# samples $=3600$} \\
\hline
\Centerstack[c]{\textbf{Reference results using the} \\ \textbf{Weibull approach from \cite{Dhulipala_TRISO}}} & \Centerstack[c]{{$\hat{P}_f=0.283$} \\ $\hat{\delta}_{P_f}=0.027$ \\ {$\hat{\beta} = 0.573$}} & \Centerstack[c]{{$\hat{P}_f= 0.12$} \\ $\hat{\delta}_{P_f}=0.036$ \\ {$\hat{\beta} = 1.17$}} \\
\hline
\end{tabular}
\centering
\begin{tablenotes}
\item[] \textbf{Notations.} $\hat{P}_f$: Failure probability estimate of the IPyC layer; $\hat{\delta}_{{P}_f}$: COV estimate over $\hat{P}_f$; $\hat{\beta}$: Reliability index estimate
\end{tablenotes}
\end{table}


Next, we computed the SiC failure conditioned on IPyC failure probabilities using the coupled active learning, multifidelity modeling, and subset simulation with the two multifidelity modeling strategies presented in Figure \ref{2DTRISO_cases}. We used four and three subsets for models 1 and 2, respectively, with 5,000 samples per subset such that the COV estimate on SiC conditional on IPyC failure probability is around 0.075. This COV value will result in an overall COV of around 0.1 or less when the IPyC failure probability is also considered. Table \ref{Table_2D_SiC_failure} presents the overall SiC (and TRISO) failure results by also considering the IPyC failure results from Table \ref{Table_2D_IPyC_failure} through Equation \eqref{eqn:TRISO_Pf}. The failure probability and reliability index estimates from the two multifidelity modeling strategies match well with the reference results using a Weibull approach from \citet{Dhulipala_TRISO} across the two models. Interestingly, Strategy (b) in Figure \ref{2DTRISO_cases}, which uses a physics-based LF model, requires about $13\%$ fewer calls on an average to the 2D TRISO model compared to Strategy (a), which uses a data-driven LF model. This is expected since the DNN in Strategy (a), which predicts the outputs in a black-box fashion is trained only on 12 evaluations of the 2D TRISO model. Contrarily, the 1D TRISO model in Strategy (b) is physics-based, and it approximates the behavior of the 2D TRISO model at a more fundamental level. {With respect to the results for the 1D TRISO Models 1 and 2 in Table \ref{Table_SiC_failure}, the failure probabilities presented below in Table \ref{Table_2D_SiC_failure} are higher. This disparity between the 1D and 2D TRISO model results is because the 1D model approximates the stress states in the 2D model through stress modification factors, which can be inconsistent with the 2D TRISO model, where the stress concentrations due to IPyC cracking are explicitly simulated using XFEM \cite{Jiang2020a}. The stress modification factors used in the 1D model could be re-calibrated by running several simulations of the both the 1D and 2D models. These manually re-calibrated stress modification factors may lead to a better agreement between the 1D and 2D TRISO failure probability results, but direct 2D model evaluations are expected to provide more accurate results.}

\begin{table}[H]
\centering
\caption{2D TRISO model overall SiC (and TRISO fuel) failure probabilities. Results corresponding to the two multifidelity modeling strategies in Figure \ref{2DTRISO_cases} are presented.}
\label{Table_2D_SiC_failure}
\small
\begin{tabular}{ |c|c|c| }
\hline
\multicolumn{3}{|c|}{\textbf{Overall SiC layer (and TRISO) failure probability for the 2D TRISO models}}\\
\hline
& \textbf{Model 1} & \textbf{Model 2}\\
\hline
\Centerstack[c]{\textbf{AK-MCS +} \\ \textbf{Data-driven multifidelity} \\ \textbf{Strategy (a) in Figure \ref{2DTRISO_cases}}} & \Centerstack[c]{$\hat{P}_f=1.62E-4$ \\ $\hat{\delta}_{P_f}=0.085$ \\ $\hat{\beta} = 3.595$ \\ \# HF calls $=212$ \\ \# samples $=21500$} & \Centerstack[c]{$\hat{P}_f=1.17E-4$ \\ $\hat{\delta}_{P_f}=0.085$ \\ $\hat{\beta} = 3.68$ \\ \# HF calls $=178$ \\ \# samples $=18600$}\\
\hline
\Centerstack[c]{\textbf{AK-MCS +} \\ \textbf{Physics-based multifidelity} \\ \textbf{Strategy (b) in Figure \ref{2DTRISO_cases}}} & \Centerstack[c]{$\hat{P}_f=1.88E-4$ \\ $\hat{\delta}_{P_f}=0.085$ \\ $\hat{\beta} = 3.56$ \\ \# HF calls $=190$ \\ \# samples $=21500$} & \Centerstack[c]{$\hat{P}_f=1.24E-4$ \\ $\hat{\delta}_{P_f}=0.086$ \\ $\hat{\beta} = 3.65$ \\ \# HF calls $=150$ \\ \# samples $=18600$}\\
\hline
\Centerstack[c]{\textbf{Reference results using the} \\ \textbf{Weibull approach from \cite{Dhulipala_TRISO}}} & \Centerstack[c]{{$\hat{P}_f=1.92E-4$}  \\ $\hat{\delta}_{P_f}=0.076$ \\ {$\hat{\beta} = 3.55$}} & \Centerstack[c]{{$\hat{P}_f=1.05E-4$}  \\ $\hat{\delta}_{P_f}=0.094$ \\ {$\hat{\beta} = 3.7$}} \\
\hline
\end{tabular}
\centering
\begin{tablenotes}
\item[] \textbf{Notations.} $\hat{P}_f$: Failure probability estimate of TRISO; $\hat{\delta}_{{P}_f}$: COV estimate over $\hat{P}_f$; $\hat{\beta}$: Reliability index estimate
\end{tablenotes}
\end{table}

\subsection{Discussion on the Computational Budget for Data-Driven and Physics-Based Low-Fidelity Models}

\begin{figure}[H]
\begin{subfigure}{0.5\textwidth}
\centering
\includegraphics[width=2.75in, height=2.5in]{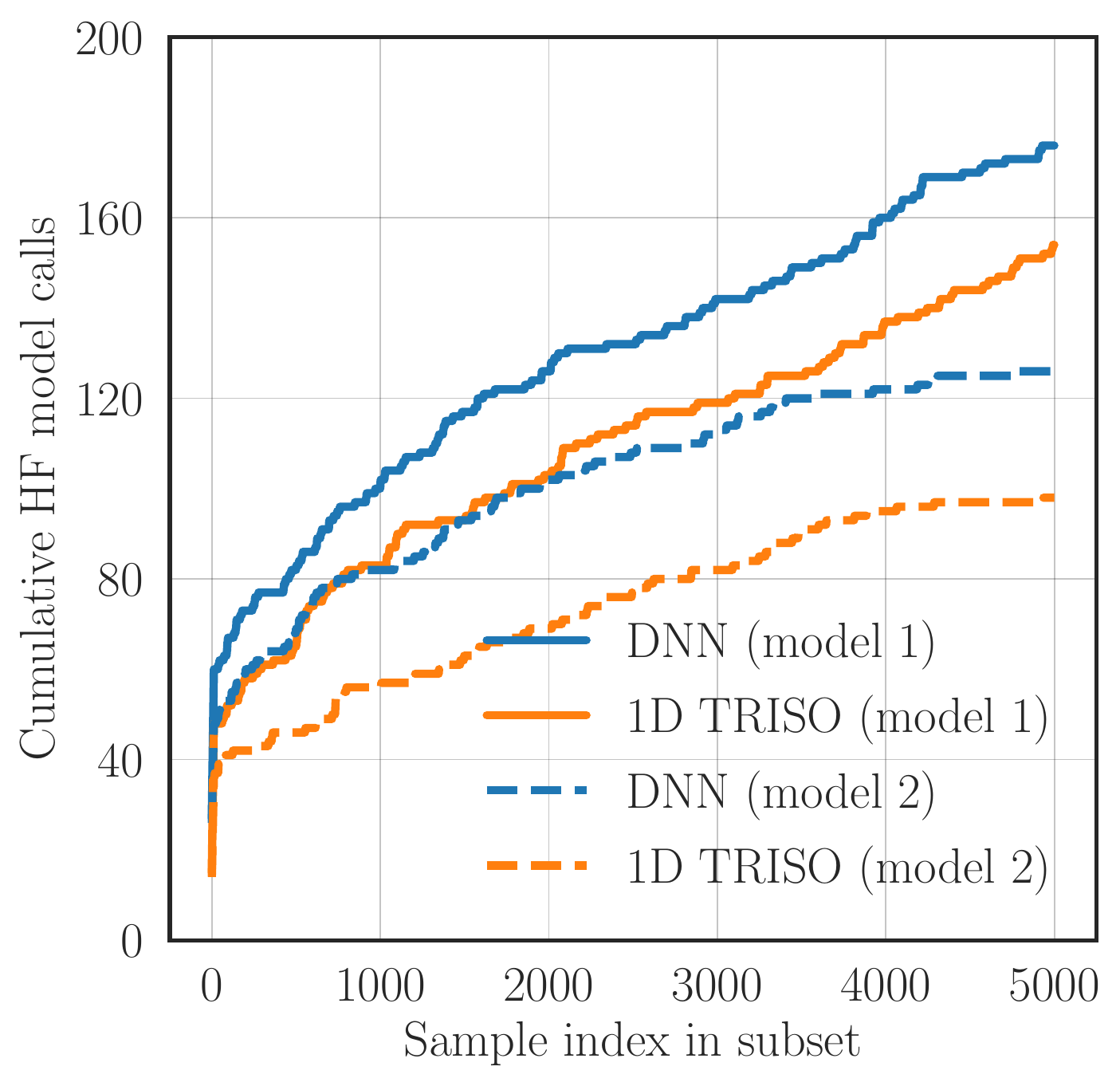}
\caption{}
\label{2DTRISO_calls}
\end{subfigure}
\begin{subfigure}{0.5\textwidth}
\centering
\includegraphics[width=2.75in, height=2.5in]{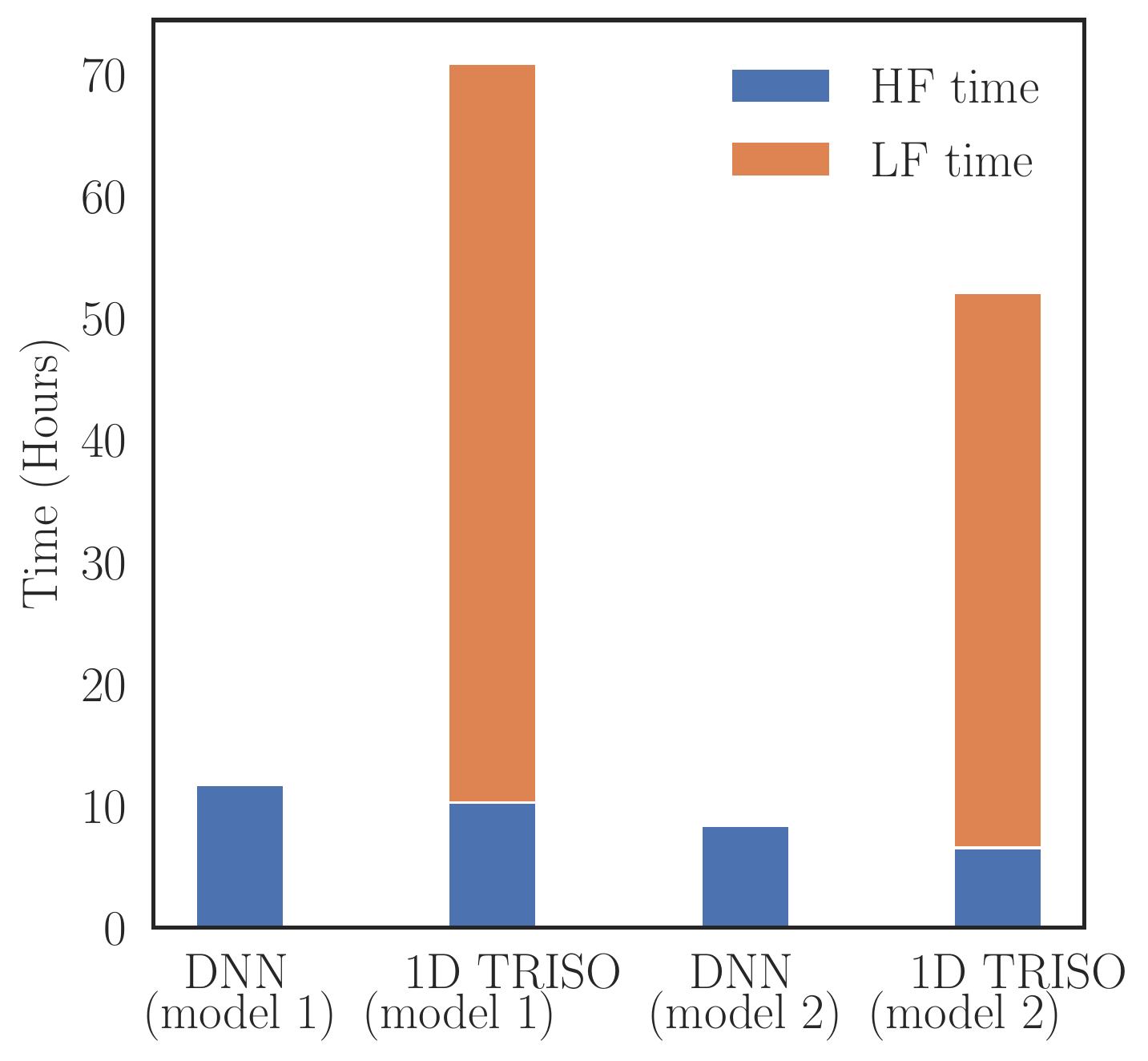}
\caption{}
\label{2DTRISO_time}
\end{subfigure}
\caption{(a) Evolution of the cumulative number of HF model (2D TRISO model) calls with the sample index in the subset for the three multifidelity modeling strategies in Figure \ref{2DTRISO_cases}. Since there are three to four subsets, at each sample index, the number of HF model calls across these subsets are added. (b) Computational budget for the two multifidelity modeling strategies in Figure \ref{2DTRISO_cases}. LF and HF here stand for low-fidelity and high-fidelity, respectively. In computing the computational budgets, we did not consider the small overhead time it takes to re-train the Kriging correction model every time a HF called is made in both the multifidelity modeling strategies (i.e., DNN or 1D TRISO as the LF model).}
\label{2DTRISO_calls_time}
\end{figure}

Figure \ref{2DTRISO_calls} plots the cumulative number of 2D TRISO model calls as a function of the sample index for the two multifidelity modeling strategies in Figure \ref{2DTRISO_cases} for the two TRISO models. At each sample index, the number of 2D TRISO calls across all the subsets are added. Only the results corresponding to the SiC failure conditioned on IPyC failure are considered here. Strategy (b) with 1D TRISO as the LF model is consistently calling the 2D TRISO model less often than Strategy (a), which uses a DNN as the LF model. However, as seen in Figure \ref{2DTRISO_time}, calling the 2D TRISO model fewer times does not necessarily result in a lower overall simulation time for estimating the failure probability and reliability index. In Strategy (a), the DNN predictions are almost instantaneous and do not contribute to the overall simulation time. However, each 1D TRISO model evaluation in Strategy (b) requires about 11 seconds and can contribute significantly to the overall simulation time. Therefore, the accuracy losses due to using a black-box DNN as the LF model are outweighed by its nearly instantaneous predictions, which leads to a significantly lower overall simulation time compared to using 1D TRISO as the LF model. {In other words, the gains in accuracy provided by using a 1D TRISO model as the LF model relative to a DNN are not enough to justify its simulation cost. For more general applications, this overall cost comparison obviously depends on the ratio of computational expense between the HF and LF models. A potential strategy to further increase computational gains could be to consider hierarchies of LF models based on their computational cost and accuracy; for example, in the TRISO case, a DNN could be the lowest fidelity model, the 1D TRISO model could be a LF model that is more accurate, but also more computationally expensive than the DNN, and the 2D TRISO model could be a HF model. The 1D TRISO model could then only be called when the active learning functions deem that the DNN prediction is inadequate, and further, the 2D TRISO model could only be called when the 1D TRISO model predictions are inadequate. In the future, we will explore Bayesian model averaging strategies to consider such hierarchies of LF models in the multifidelity modeling framework for reliability estimation.}

\section{Summary and Conclusions}\label{sec:sum_conc}

TRISO nuclear fuel is a robust fuel that is being proposed for use in several advanced reactor technologies, such as small modular reactors and microreactors. Estimating the failure probability and reliability of TRISO fuel is critical for the success of these technologies since these calculations further permit analyzing fission product release into the environment and assessing reactor safety. However, the failure probability estimates of TRISO fuel are usually small, from $10^{-3}$ to $10^{-6}$, and the computational models for TRISO can be expensive. To this end, we applied a coupled active learning, multifidelity modeling, and subset simulation algorithm to efficiently estimate the reliability of TRISO fuel. This algorithm has several advantages such as: (1) the use of corrected LF models to better approximate the HF model output in a multifidelity strategy, unlike the single surrogate model strategy used in most active learning methods; (2) sequential active learning where the test size for Kriging predictions is always one; and {(3) the use of subset-dependent active learning functions, which allow the algorithm to even estimate very small failure probabilities of the order $10^{-9}$.} Sequential active learning means that the algorithm does not require Kriging evaluations on $N_o$ Monte Carlo samples used in most active learning methods, which may cause computational and memory bottlenecks for smaller failure probabilities.

We first applied the coupled active learning algorithm to four 1D TRISO models (requiring $\sim11$ seconds for each evaluation) using three strategies for multifidelity modeling: (a) Kriging as an actively learned surrogate model; (b) Kriging as an LF model plus an added Kriging correction term; and (c) DNN as an LF model plus an added Kriging correction term. In all three strategies, the 1D TRISO model was used as the HF model. Across all four 1D TRISO models, all three strategies satisfactorily predicted the failure probabilities and reliability indices. However, Strategy (c), which uses a DNN, consistently required the fewest calls to the HF model, about $26\%$ and $18\%$ fewer compared to Strategies (a) and (b), respectively. We hypothesize that better regularization of the loss function in the DNN avoided over-fitting the training dataset leading to a better performance than Strategies (a) and (b).

Next, we applied the coupled active learning algorithm to two 2D TRISO models (requiring $\sim4$ minutes for each evaluation on 10 processors) using two strategies for multifidelity modeling: (a) DNN as an LF model plus an added Kriging correction term; and (b) 1D TRISO as an LF model plus an added Kriging correction term. Strategies (a) and (b) are data-driven and physics-based, respectively. Both the strategies satisfactorily predicted the failure probabilities and reliability indices. Although Strategy (b) required {$13\%$} fewer calls to the HF model, its overall simulation time is considerably higher than Strategy (a) since the 1D TRISO model itself has a non-negligible simulation time. In contrast, the DNN predictions are nearly instantaneous and the computational losses due to using this black-box DNN as the LF model are outweighed by its instantaneous predictions, which led to a significantly lower overall simulation time compared to using 1D TRISO model as the LF model. For more general applications, this overall cost comparison obviously depends on the ratio of computational expense between the HF and LF models. Future work will explore the application of Bayesian model averaging strategies to consider hierarchies of LF models for reliability estimation.

\section*{Acknowledgments}

We thank Peter German at Idaho National Laboratory for his valuable feedback on a draft of this paper.

This manuscript has been authored by Battelle Energy Alliance, LLC under Contract No.~DE-AC07-05ID14517 with the US Department of Energy. The United States Government retains and the publisher, by accepting the article for publication, acknowledges that the United States Government retains a nonexclusive, paid-up, irrevocable, worldwide license to publish or reproduce the published form of this manuscript, or allow others to do so, for United States Government purposes.

This research is supported through the INL Laboratory Directed Research \& Development (LDRD) Program under DOE Idaho Operations Office Contract DE-AC07-05ID14517. This research made use of the resources of the High Performance Computing Center at INL, which is supported by the Office of Nuclear Energy of the U.S. DOE and the Nuclear Science User Facilities under Contract No. DE-AC07-05ID14517.

\section*{Appendix A: Coupled Kriging and Subset Simulation to Estimate a Very Low Failure Probability of the Order $10^{-9}$}\label{section_app}

We use a borehole reliability problem \cite{Schobi2017a} to demonstrate that the coupled Kriging and subset simulation method presented in Figure \ref{SS_GP} can estimate a very low failure probability of the order $10^{-9}$. For this demonstration, we do not use the multifidelity modeling concept presented in Section \ref{sec:MFM}. The borehole function is defined as:

\begin{equation}
    \label{Borehole_1}
    F(\pmb{X}) = \frac{2\pi~T_u~(H_u-H_l)}{\ln{(r/r_w)}~\bigg(1+\frac{2LT_u}{\ln{(r/r_w)}~r_w^2~K_w}+\frac{T_u}{T_l}\bigg)}
\end{equation}

\noindent where $F(\pmb{X})$ is the water flow and $\pmb{X} = \{r_w,~r,~T_u,~H_u,~T_l,~H_l,~L,~K_w\}$ is the input parameter vector. A description of $\pmb{X}$ and the distributions of the parameters are provided in \citet{Schobi2017a}. The indicator function for model failure is defined as:

\begin{equation}
    \label{Borehole_2}
    \mathbf{I} = \begin{cases}
    1~~~~~\textrm{if}~~F(\pmb{X})-300.0 \geq 0\\
    0~~~~~\textrm{if}~~F(\pmb{X})-300.0 < 0
    \end{cases}
\end{equation}

\noindent To estimate the failure probability, we ran the coupled Kriging and subset simulation method with eight subsets and $50,000$ samples per subset. We used a new Kriging model for each subset by training it on 12 model evaluations at the beginning of the subset. Although the use of a new Kriging model for each subset is unnecessary, when failure probabilities are very low, doing so may lead to computational gains since the training cost for Kriging increases cubically with the training set size. Figures \ref{App_Borehole_1} and \ref{App_Borehole_2} present the model value and cumulative number of model calls, respectively, (i.e., Equation \eqref{Borehole_1}) with respect to the sample index in each subset. Table \ref{Table:Borehole} presents the failure probability, COV, reliability index, and number of model calls. Also presented in Table \ref{Table:Borehole} are the reference results from two previous studies that have used extended AK-MCS \cite{Razaaly2020a} and importance sampling with Gaussian mixture sampling density \cite{Razaaly2018a} for this problem. It is observed that for a COV of 0.038, the failure probability and reliability index estimates provided by the coupled Kriging and subset simulation method are quite close to the reference results.

\renewcommand{\thefigure}{A1}
\begin{figure}[h]
\begin{subfigure}{0.5\textwidth}
\centering
\includegraphics[width=2.5in, height=2.25in]{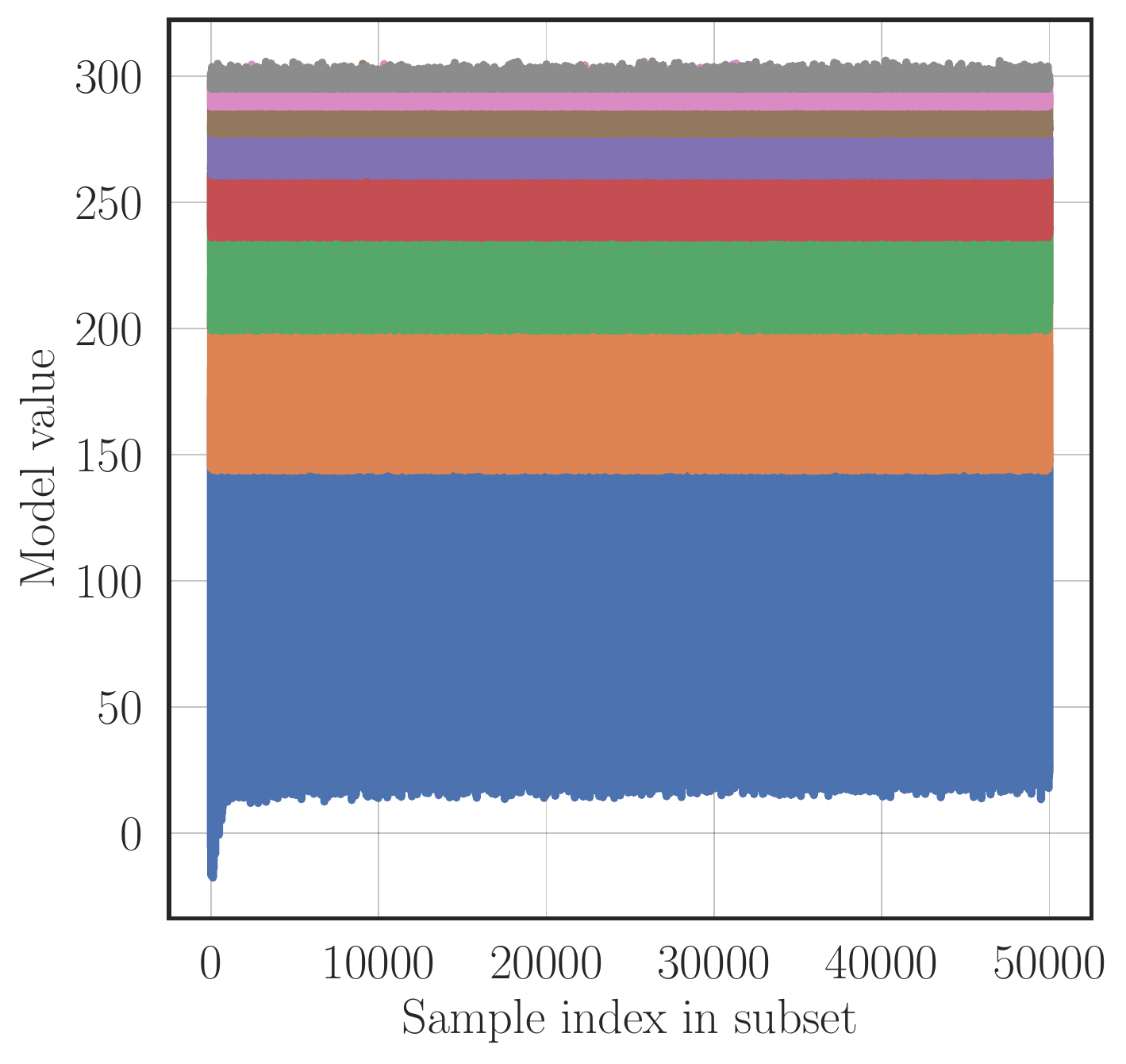}
\caption{}
\label{App_Borehole_1}
\end{subfigure}
\begin{subfigure}{0.5\textwidth}
\centering
\includegraphics[width=2.5in, height=2.25in]{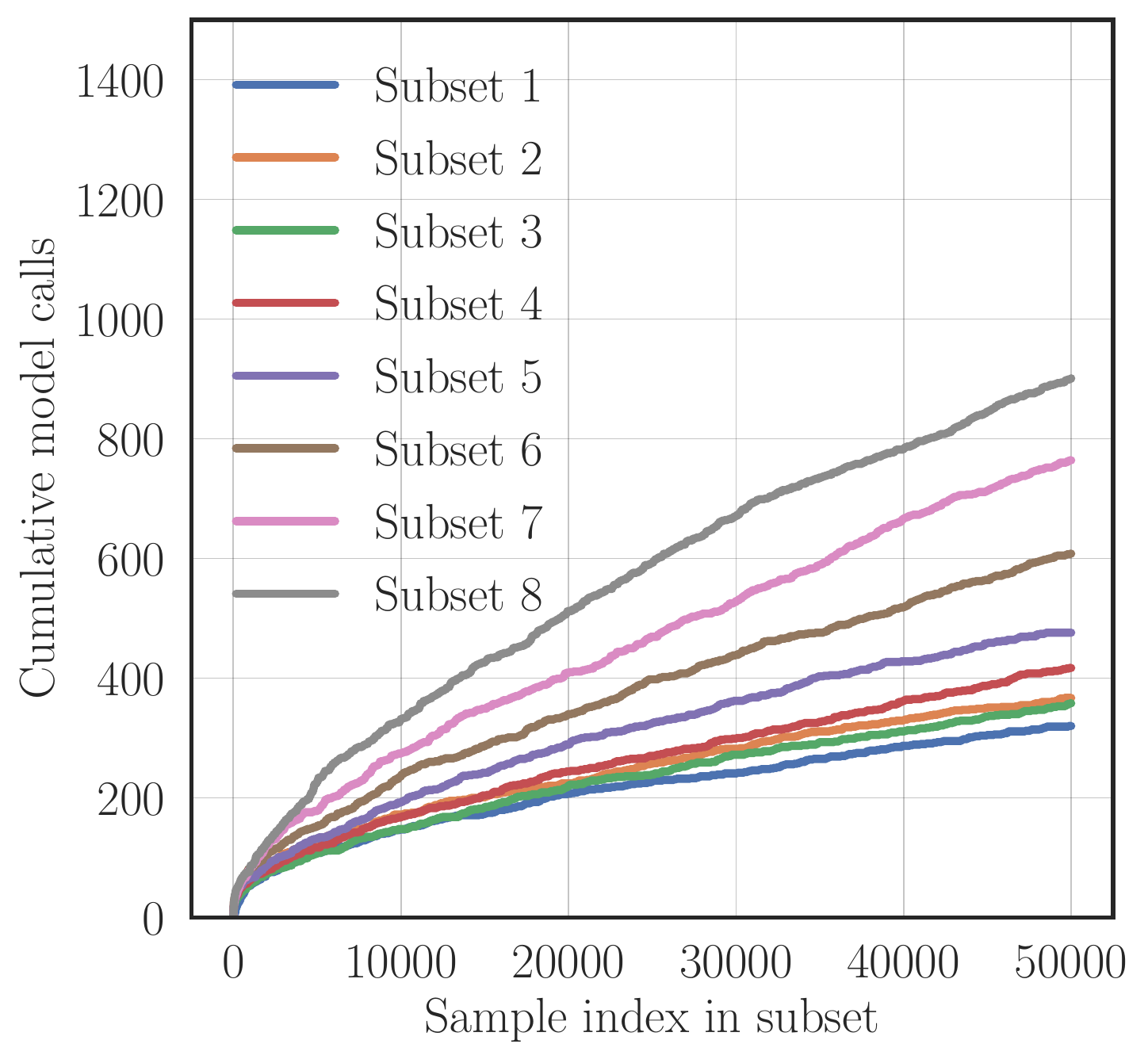}
\caption{}
\label{App_Borehole_2}
\end{subfigure}
\caption{(a) Borehole model value evolution within each subset as well across subsets in the coupled Kriging and subset simulation algorithm in Figure \ref{SS_GP}. (b) Cumulative number of borehole model calls with the sample index in each subset.}
\label{App_Borehole}
\end{figure}

\renewcommand{\thetable}{A1}
\begin{table}[h]
\centering
\caption{Borehole model failure probability and reliability estimated using the coupled Kriging and subset simulation algorithm in Figure \ref{SS_GP}. Reference results from \citet{Razaaly2020a} and \citet{Razaaly2018a} are also provided for comparison.}
\label{Table:Borehole}
\small
\begin{tabular}{ |c|c|c|c|c| }
\hline
\textbf{Method} & \Centerstack[c]{\textbf{Failure}\\\textbf{Probability}~$P_f$} & \Centerstack[c]{\textbf{COV}\\\textbf{for}~$P_f$} & \Centerstack[c]{\textbf{Reliability}\\\textbf{Index}} & \Centerstack[c]{\textbf{Model}\\\textbf{Calls}}\\
\hline
\Centerstack[c]{Coupled Kriging\\and subset simulation\\(Figure \ref{SS_GP})} & $8.45E-9$ & $0.038$ & $5.64$ & $4211$ \\
\hline
eAK-MCS \cite{Razaaly2020a} & $8.97E-9$ & 0.002 & $5.63$ & --\\
\hline
\Centerstack[c]{Importance sampling \\ with Gaussian mixture\\ sampling density \cite{Razaaly2018a}} & $8.73E-9$ & $0.003$ & $5.64$ & $1E7$\\
\hline
\end{tabular}
\end{table}

\bibliography{main}

\end{document}